\documentclass[trackchanges,twocolumn]{aastex701}

\usepackage{amsmath}
\usepackage{appendix}

\newcommand{\diff}{\ensuremath{\rm d}}

\begin{document}


\title{Theoretical Constraints on Neutron Star Superfluidity from Her X-1 Precession}

\author[orcid=0000-0001-7210-6988,sname='Biryukov']{Anton Biryukov}
\affiliation{School of Physics and Astronomy, Tel Aviv University, Tel Aviv, 6997801, Israel}
\affiliation{Sternberg Astronomical Institute, Lomonosov Moscow State University, 13 Universitetsky pr., Moscow, 119234, Russia}
\email[show]{ant.biryukov@gmail.com}  

\author[sname='Levinson', orcid=0000-0001-7572-4060]{Amir Levinson}
\affiliation{School of Physics and Astronomy, Tel Aviv University, Tel Aviv, 6997801, Israel}
\email[show]{levinson@tauex.tau.ac.il}

\author[sname='Abolmasov', orcid=0000-0002-6429-0436]{Pavel Abolmasov} 
\affiliation{School of Physics and Astronomy, Tel Aviv University, Tel Aviv, 6997801, Israel}
\email{}

\begin{abstract}
Recent IXPE observations of Her X-1 reveal correlations between flux, polarization degree, and polarization angle with its 35-day superorbital cycle. These measurements have been interpreted as strong evidence that the 35-day period is driven by nearly free precession of the neutron star. We show that this interpretation carries far-reaching implications for the dynamics of the crustal superfluid. In particular, maintaining precession over the $\sim 50$-year observational baseline of Her X-1 would require that superfluid vortices remain unpinned for centuries and experience extremely weak mutual friction while traversing the heavy-ion lattice of the inner crust -- conditions that challenge conventional wisdom and standard models of glitch dynamics.
\end{abstract}


\section{Introduction}
\label{sec:intro}
Neutron stars are born hot but cool rapidly, and a substantial fraction of their interior is expected to transition into a superfluid state within years to decades after formation. In particular, neutrons in the inner crust and outer core are widely believed to form a superfluid capable of sustaining long-lived, weakly dissipative flows on macroscopic scales. In a rotating neutron star, the angular momentum of this superfluid is carried by an array of quantized vortex lines \citep{baym69}, whose areal density is directly proportional to the coarse-grained angular velocity of the superfluid \citep{khalatnikov1965,Baym1983}.

It has long been thought that pulsar glitches -- sudden discrete increases in the observed rotation frequency -- provide some of the most direct evidence for the presence of a superfluid component in neutron stars \citep[for reviews, see][and references therein]{Haskell15,antonelli2022}. In the standard picture, vortices in the crustal superfluid are partially pinned to the nuclear lattice, allowing the superfluid to rotate slightly faster than the solid crust as the star spins down under electromagnetic torques. Differential rotation then builds up until a critical threshold is reached, triggering rapid vortex unpinning and outward motion \citep{anderson1975,alpar1984a,alpar1984b,link2014}. The resulting transfer of angular momentum from the superfluid to the crust produces the observed glitch.

Following a glitch, the star typically exhibits a relaxation phase, during which the rotation rate evolves toward a new equilibrium on timescales ranging from days to months. This post-glitch recovery is commonly interpreted as the gradual re-coupling of superfluid components that were only weakly coupled to the crust during the glitch itself. In particular, mutual interactions between vortices and the surrounding charged fluid mediate the exchange of angular momentum between the crust and the superfluid in the core, leading to partial or complete recovery of the pre-glitch spin-down rate. The detailed form of the relaxation thus encodes information about the strength of superfluid coupling and the distribution of superfluid reservoirs within the star.

Vortex pinning can have a profound effect on the free precession of a neutron star by dynamically coupling the superfluid component to the solid crust. In the inner crust, quantized vortices of the neutron superfluid are expected to pin to the nuclear lattice, preventing the superfluid from freely adjusting its angular momentum in response to changes in the crust’s rotation. If vortices remain strongly pinned during precession, the pinned superfluid exerts a dominant gyroscopic restoring force on the crust that controls the crust’s precessional dynamics. As originally shown by \citet{Shaham77} (see also \citealt{jones2001,link2002}), in case of absolute pinning the precessional angular frequency must be of the order $\omega_{\rm p} \sim (I_{\rm s}/I)\Omega_\mathrm{s} \gtrsim 10^{-2}\Omega$, where $I_{\rm s}/I$ is the fraction of the stellar moment of inertia in the pinned superfluid and $\Omega$ is the angular frequency of the star.

The fast precessional periods (a few minutes typically) imposed by absolute pinning are inconsistent with the much longer precessional periods inferred for some neutron stars, most notably Hercules X-1. It has long been suspected that the observed 35-day superorbital period in Hercules X-1 may be due to (nearly free) precession of the neutron star. Recent X-ray polarization measurements with the Imaging X-ray Polarimetry Explorer (IXPE), which can directly probe the spin geometry of the neutron star, have recently been argued to support this interpretation \citep{Heyl24,Zhao2024}. 

\citet{Alpar87} proposed that a steady state may exist in which a pinned superfluid follows the precessional motion of the crust through thermal vortex creep. In this picture, the component of the superfluid–crust lag perpendicular to the vortex lines is of order the precession frequency $\omega_{\rm p}$ and therefore much smaller than the critical lag required for unpinning. The relaxation time to this equilibrium depends on the ratio of the pinning energy to the thermal energy and was estimated to be of the order of a few hundred to a few thousand years for Her~X--1. However, \citet{Alpar87} neglected the non-dissipative torque associated with the component of the mutual-friction force acting in the direction $\pmb{\tilde\omega}\times(\pmb{v}_{\rm n}-\pmb{v}_{\rm s})$, where $\pmb{v}_{\rm n}$ and $\pmb{v}_{\rm s}$ are the normal-fluid and superfluid velocities, respectively, and $\pmb{\tilde\omega}$ is the superfluid vorticity.  As we shall show, this omission is not justified.

The torque associated with the mutual-friction force was derived rigorously by \citet{Sedrakian1999} and applied to the calculation of precessional modes in the case of imperfect pinning. In the strong-drag regime, they found low-frequency, strongly damped free-precession modes and concluded that
their results are qualitatively consistent with \cite{Shaham77}, in that no long-lived free precession survives in the strong-drag regime. However, their analysis is perturbative and restricted to oblate deformations, and considers only small (linear) lags about a fixed point in which all angular velocities are equal and aligned with a stable principal axis. It therefore remains unclear how the rotational dynamics behaves in the general case.

In this paper, we present a comprehensive analysis of the rotational evolution, taking into account the internal torque exerted on the normal (rigid) crust by the crustal superfluid and assuming tight coupling between the core and the normal crust. We apply these results to Her X-1 and show that if the 35-d cycle is produced by nearly free precession of the neutron star, then superfluid vortices must remain unpinned for centuries or longer and experience extremely weak mutual friction, with a dimensionless drag coefficient $\lesssim 0.01$, while traversing the heavy-ion lattice of the inner crust. This implies that the existence of long-lived precession is incompatible with the buildup of the large differential rotation required to trigger glitches. 

\section{Two-component Model}
\label{sect:model}

Three main components govern a neutron star’s rotational dynamics: the outer core, the elastic charged lattice in the crust, and the crustal neutron superfluid. In isolated neutron stars, the core temperature is expected to fall below the neutron superfluid transition temperature, $T_{\rm n} \sim 5\times 10^8$ K, so that neutrons in the core become superfluid. In accreting neutron stars, however, the core temperature may approach $T_{\rm n}$ (\citealt{brown2000}; but see \citealt{nava-c2025}), allowing neutrons in much of the outer core to remain in a normal state (Wasserman, private communication). In both cases, the temperature is well below the critical temperature for proton pairing, so that protons form either a Type I or Type II superconductor. \cite{link2003} argued that if the core contains coexisting superfluid vortices and superconducting proton flux tubes, pinning between vortices and flux tubes would suppress long-period precession, unless vortices can somehow cut through the flux tubes. Here, we ignore pinning of vortices to flux tubes
(as expected, e.g., when the neutrons in the core are normal or the protons are in an intermediate state). Under these conditions, the core–crust coupling is mediated by electromagnetic forces, with a strength that depends on the detailed microphysics (see, e.g., \citealt{Levin2004} for detailes). Here, we adopt, for illustration, the simple phenomenological model of \cite{Alpar87} in which the torque is given by:
\begin{equation}
    \pmb N_\mathrm{co} = I_\mathrm{co}\tau_\mathrm{co}^{-1}(\pmb \Omega_\mathrm{co} - \pmb\Omega_\mathrm{n}),
\end{equation}
where the relaxation time $\tau_\mathrm{co}$ depends on the specific coupling mechanism.
Here $I_\mathrm{co}$ and $\pmb\Omega_\mathrm{co}$ are, respectively, the moment of inertia and angular velocity vector of the core, 
and $\pmb\Omega_\mathrm{n}$ is the angular velocity vector of the elastic (normal) crust.  In the case of electron scattering off magnetized superfluid 
vortices \cite{Alpar1988} obtained 
\begin{equation}
   \tau_\mathrm{co} \sim 400-10^4 P_\mathrm{n}
   \label{eq:tau_d_core}
\end{equation}
where $P_\mathrm{n} = 2\pi/\Omega_\mathrm{n}$ is the rotation period, about 1.2 sec in Her X-1.  In case where the neutrons in the core are normal, the coupling time depends on the strength and geometry of the magnetic field in the core. 
In Appendix \ref{app:core_effect}, we show that, for a deformed crust with ellipticity $\epsilon_\mathrm{cr}$ and an isotropic core, free precession is damped on a timescale
\begin{equation}
   \tau_\mathrm{p} \approx \frac{I_\mathrm{n}^2(\Omega_\mathrm{n} \tau_\mathrm{co})^2 +I_\mathrm{co}^2}{\epsilon_\mathrm{cr} \, I_\mathrm{co} I_\mathrm{n} \Omega_\mathrm{n}^2\tau_\mathrm{co}},
   \label{eq:tau_p_core}
\end{equation}
where $I_\mathrm{co}$ and $I_\mathrm{n}$ are the moments of inertia of the core and crust, respectively. 
As expected, $\tau_\mathrm{p}\to \infty$ both in the limit $\tau_\mathrm{co} \to \infty$ (no coupling) and $\tau_\mathrm{co}\to 0$ (infinitely tight coupling).
Using Eq. \eqref{eq:tau_d_core}, we find that $\Omega_\mathrm{n} \tau_\mathrm{d} \gg I_\mathrm{co}/I_\mathrm{n}$ for a fiducial ratio $I_\mathrm{co}/I_\mathrm{n}\approx 10^2$.
In this regime
\begin{equation}
\tau_\mathrm{p}\approx \epsilon_\mathrm{cr}^{-1}\left(\frac{I_\mathrm{n}}{I_\mathrm{co}}\right)\, \tau_\mathrm{co}  \approx 3\times10^{2}\,\tau_\mathrm{co} ,
\end{equation}
adopting $\epsilon_\mathrm{cr} = (\Omega_\mathrm{p}/\Omega_\mathrm{n})(I_\mathrm{co}/I_\mathrm{n}) = 3\times10^{-5}$ for a precession frequency of $\Omega_\mathrm{p} = 2\pi/35\, \text{days}$, 
as inferred from observations of Her X-1.
For $\tau_\mathrm{co} \sim 10^{4} P_\mathrm{n}$, this gives $\tau_\mathrm{p} \sim 0.1\,\mathrm{yr}$, shorter than the observational span ($\sim 50$ yr). On the other hand, a deformed core will naturally induce deformations in the overlying crust. During precession, the crust continuously slips against the precessing core, a process expected to cause precession damping. A detailed analysis of the full core–crust interaction is beyond the scope of this work. Here, to obtain a rough estimate of the damping timescale, we assume an isotropic moment of inertia for the crust. Under this simplifying assumption one finds
\begin{equation}
   \tau_\mathrm{p} \approx \epsilon_\mathrm{co}^{-1}\left(\frac{I_\mathrm{co}}{I_\mathrm{n}}\right)\tau_\mathrm{co}
   \approx 3\times10^{8}\,\tau_\mathrm{co} \approx 10^5 \, \text{yr},
\end{equation}
here $\epsilon_\mathrm{co} = \Omega_\mathrm{p}/\Omega_\mathrm{n} \approx 3\times10^{-7}$ is the ellipticity of the core, indicating that the core and solid crust can be treated as a single rigid body on observational timescales (see also \citealt{Levin2004}).

The deformation of the core may be caused by magnetic stresses. If the protons in the core form a type-II superconductor, the resulting ellipticity can be written as \citep{cutler2002}
\begin{equation}
 \epsilon_\mathrm{co} = \frac{25r_\mathrm{NS}^4}{24GM_\mathrm{NS}^2}\left(2\langle B_\mathrm{p} H_\mathrm{c1}\rangle - \langle B_\mathrm{t} H_\mathrm{c1}\rangle\right),
\end{equation}
for $B_\mathrm{p},B_\mathrm{t} \ll H_\mathrm{c1}$, where $B_\mathrm{p}$ and $B_\mathrm{t}$ are the poloidal and toroidal magnetic-field components in the core, angle brackets mean volume averaging over the interior, and
$H_\mathrm{c1}\approx 10^{15}$ G is the lower critical field for the superconducting transition. 
Thus, a toroidal field $B_\mathrm{t} \sim 30 B_\mathrm{p} \approx 10^{14}$ G is sufficient to produce the required ellipticity, assuming $B_\mathrm{p}\approx 4\times 10^{12}$ G -- 
roughly the measured poloidal field near the stellar surface in Her X-1.  In this case pinning  between vortices and flux tubes 
may be avoided if the neutrons are normal.

To simplify the analysis, we treat the core and the elastic crust as a single deformed rigid body. 
This is an optimal-case assumption, since any friction between the core and the crust would only increase the damping of precession.
A fully self-consistent treatment incorporating all three components is deferred to future work.

\subsection{Mutual-friction torques in the crust}
\label{sec:model:friction}
The coupling between the crustal superfluid and the elastic crust is mediated by interactions between quantized vortices and crustal 
nuclei. This interaction is commonly modeled by a phenomenological drag force per unit length of a vortex: $\pmb f_\mathrm{d} = -
\eta (\pmb v_\mathrm{L} -\pmb v_\mathrm{n})$, where $\eta$ is a drag coefficient, $\pmb v_\mathrm{L}$ is the vortex velocity and $\pmb 
v_\mathrm{n} = \pmb \Omega_\mathrm{n} \times \pmb r$ is the velocity of the normal crustal component. 
The motion of a vortex is determined by balancing this drag force with the Magnus force arising from the relative motion between the bulk superfluid and the vortex line, 
$\pmb f_\mathrm{M} = \rho_\mathrm{s} \kappa \, \hat{\pmb \omega} \times (\pmb v_\mathrm{L} - \pmb v_\mathrm{s})$, which leads to
\begin{equation}
    \rho_\mathrm{s} \kappa \, \hat{\pmb \omega} \times (\pmb v_\mathrm{L} - \pmb v_\mathrm{s}) = \eta (\pmb v_\mathrm{L} -\pmb v_\mathrm{n}).
\end{equation}
Here $\kappa = h/2m_\mathrm{n}$ is the quantum of circulation, $\hat{\pmb \omega}$ is a unit vector in the direction of local vorticity,   
$\rho_s$ is the superfluid mass density, and $\pmb v_\mathrm{s}$ is its local velocity.  The latter equation can be solved to yield 
\begin{equation}
\begin{split}
    \pmb v_\mathrm{L} - \pmb v_\mathrm{n} & = \frac{{\cal R}}{1+{\cal R}^2}\hat{\pmb \omega}\times(\pmb v_\mathrm{n}-\pmb v_\mathrm{s})
    \\ &\quad + \frac{1}{1+{\cal R}^2}\hat{\pmb \omega}\times \left(\hat{\pmb \omega}\times(\pmb v_\mathrm{n}-\pmb v_\mathrm{s}) \right),
\end{split}
\end{equation}
where $\mathcal{R} =\eta/\rho_\mathrm{s}\kappa$ is the dimensionless drag coefficient. Substituting $\pmb v_\mathrm{L} -\pmb v_\mathrm{n}$ from the last equation into the drag force $\pmb{f}_\mathrm{d}$
yields the mutual friction force density:
\begin{equation}
\begin{split}
\pmb{f}_\mathrm{MF}
& = \rho_\mathrm{s} \kappa n_\mathrm{v} 
\left[
\frac{\mathcal{R}^2}{1+\mathcal{R}^2}\, \hat{\pmb \omega} \times (\pmb{v}_\mathrm{n} - \pmb{v}_\mathrm{s}) \right .
\\
& \quad +\left . \frac{\mathcal{R}}{1+\mathcal{R}^2}\, \hat{\pmb\omega} \times 
\left ( \hat{\pmb\omega} \times (\pmb{v}_\mathrm{n} - \pmb{v}_\mathrm{s}) \right ) \right ],
\end{split}
\end{equation}
where $n_\mathrm{v}$ is the vortex density.
Assuming that the angular velocity of the normal crust is uniform, the internal torque is given by \citep{Sedrakian1999}:
\begin{equation}
    \pmb{N}_s = \int d^3r \, (\pmb{r}\times \pmb{f}_\mathrm{MF})= \pmb{N}_{s1} +\pmb{N}_{s2},
\end{equation}
where
\begin{equation}
    {\pmb N}_{s1}  = \frac{\mathcal{R}}{1+\mathcal{R}^2}\ I_\mathrm{s}\left[ \Omega_\mathrm{s} \left(2 - \dfrac{\pmb\Omega_\mathrm{n} \cdot \pmb\Omega_\mathrm{s}}{\Omega^2_\mathrm{s}} \right)\pmb\Omega_\mathrm{s} - \Omega_\mathrm{s}\pmb\Omega_\mathrm{n} \right], 
 \end{equation}
 is the {\it dissipative} torque component, for which $\pmb N_{s1}\cdot(\pmb\Omega_\mathrm{n} -\pmb\Omega_\mathrm{s})<0$, and 
 \begin{equation}   
    {\pmb N}_{s2} = \frac{\mathcal{R}^2}{1+\mathcal{R}^2} I_\mathrm{s}(\pmb{\Omega}_\mathrm{s} \times \pmb{\Omega}_\mathrm{n}),
\end{equation}
is the non-dissipative torque component.  In general, the  drag force is nonlinear, specifically, the drag parameter $\mathcal{R}$ depends on the relative velocity
$\pmb v_\mathrm{L} -\pmb v_\mathrm{n}$ \citep[e.g.,][]{celora2020,sheng2025}.  However, this nonlinearity should not affect our conclusions.  Therefore, to simplify the analysis we adopt $\mathcal{R} =$ const.

\subsection{Euler's equations}
\label{sec:model:Euler}
As stated above, the two-component model adopted here treats the core and normal crust as a single deformed rigid body.  It is, therefore, convenient to 
solve Euler's equations in the frame rotating with the normal crust.  In this frame, the equations read:
\begin{align}
    \frac{\diff}{\diff t} \pmb{L}_\mathrm{n} &+ \pmb{\Omega}_\mathrm{n} \times \pmb{L}_\mathrm{n} = \pmb{N}_\mathrm{ext} + \pmb{N}_\mathrm{s}, \label{eq:Euler_basic1} \\
    \frac{\diff}{\diff t} \pmb{L}_\mathrm{s} &+ \pmb{\Omega}_\mathrm{n} \times \pmb{L}_\mathrm{s} = - \pmb{N}_\mathrm{s} , \label{eq:Euler_basic2}
\end{align}
where $\pmb{N}_\mathrm{ext}$ is the external torque acting on the crust. For simplicity, we assume the moment of inertia tensor of the crustal superfluid to be isotropic, $\pmb{I}_\mathrm{s} = I_\mathrm{s}\, \pmb{\delta}$, where $\pmb{\delta}$ is the unit tensor. It follows that $\pmb{L}_\mathrm{s} = I_\mathrm{s} \pmb{\Omega}_\mathrm{s}$.

The moment of inertia of the coupled normal crust and core is, in general, anisotropic. Following \citet{Biryukov2025}, we write its 
angular momentum in the principal-axis frame  as
$\pmb{L}_\mathrm{n} = I_0 \pmb{\Omega}_\mathrm{n} + I_0 \pmb{\omega}$. Here, $I_0$ is the moment of inertia of the crust about the intermediate principal axis, and
$\pmb{\omega} = (\epsilon_1 \Omega_1, 0, \epsilon_3 \Omega_3)^T$, with $\epsilon_1 \le 0$ and $\epsilon_3 \ge 0$.

With the above assumptions, Eqs. \eqref{eq:Euler_basic1} and \eqref{eq:Euler_basic2} yield: 
\begin{equation}
    \pmb{\dot\Omega}_\mathrm{n} + \pmb{\dot\omega} = (\pmb \omega \times \pmb \Omega_\mathrm{n}) + \dfrac{\pmb N_\mathrm{ext}}{I_0}
    + \dfrac{\pmb N_\mathrm{s1} + \pmb{N}_\mathrm{s2}}{I_0},
    \label{eq:Euler_crust}
\end{equation}
and
\begin{equation}
    \pmb{\dot\Omega}_\mathrm{s}  = (\pmb \Omega_\mathrm{s} \times \pmb\Omega_\mathrm{n})-\dfrac{\pmb N_\mathrm{s1} + \pmb N_\mathrm{s2}}{I_\mathrm{s}}.
    \label{eq:Euler_superfluid}
\end{equation}
respectively.
%

\section{Approximate analytic solutions}
\label{sec:analytic}

\begin{figure}
    \centering
    \includegraphics[width=\columnwidth]{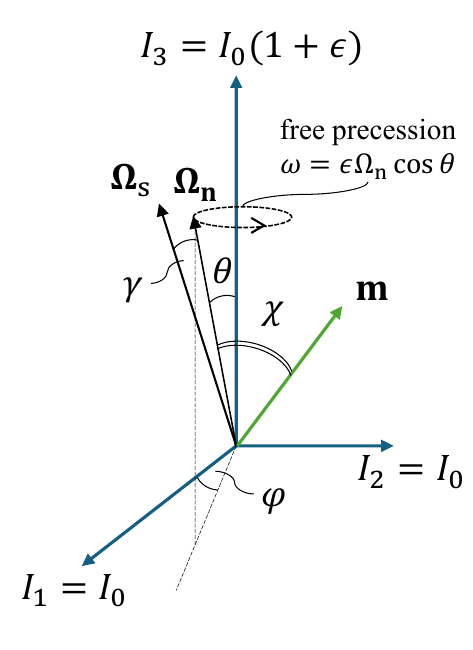}
    \caption{Reference frame aligned with the principal moments of inertia of the neutron-star normal crust, $I_1$–$I_2$–$I_3$. In this work we consider mostly a biaxial star, such that $I_1 = I_2 = I_0$, where $I_0 = I_\mathrm{n} + I_\mathrm{co}$ is the combined moment of inertia of the normal crust and the core, whose rotation is assumed to be tightly coupled. The vector $\pmb{\Omega}_\mathrm{n}$ denotes the angular velocity of the normal crust, while $\pmb{\Omega}_\mathrm{s}$ is the angular velocity of the crustal superfluid component.  The total effective deformation of the star is $\epsilon$, so that a purely rigid body would precess about the symmetry axis ($I_3$) with angular velocity $\Omega_\mathrm{p} = \epsilon \Omega_\mathrm{n} \cos\theta$. The angles $\varphi$ and $\theta$ are the spherical coordinates associated with this reference frame, with $\theta$ being the precession (wobbling) angle. The unit vector $\pmb{m}$ is aligned with the magnetic axis and is fixed within this frame. Angle $\chi$ is the angle between the magnetic and spin axes.}
    \label{fig:coords}
\end{figure}

In what follows, we consider, for simplicity, a biaxial star with ellipticity $\epsilon_3=\epsilon$ and $\epsilon_1 \equiv 0$ without loss of generality.  In the principal-axis frame, 
the vector $\pmb{\omega}$ takes the form $\pmb{\omega} = (0,0,\omega)^T$, where
\begin{equation}
    \omega = \epsilon \Omega_\mathrm{n} \cos\theta
\end{equation}
is the free-precession frequency in the absence of a superfluid component. 
Here, $\theta$ denotes the angle between $\pmb{\Omega}_\mathrm{n}$ and the symmetry axis 
(the wobble angle). The angular velocity of the normal crust can then be written in the principal frame as
\begin{equation}
    \pmb{\Omega}_\mathrm{n} = 
    \left ( \begin{array}{c}
        \Omega_\mathrm{n} \sin\theta\cos\varphi \\
        \Omega_\mathrm{n} \sin\theta\sin\varphi\\
        \Omega_\mathrm{n} \cos\theta\\
    \end{array} \right ),
    \label{eq:Omega}
\end{equation}
where $\varphi$ is the corresponding azimuthal angle (see Figure~\ref{fig:coords}).  The angular velocity of the superfluid can be expressed 
similarly by replacing $\Omega_\mathrm{n}, \theta, \varphi$ with $\Omega_\mathrm{s}, \theta_\mathrm{s}, \varphi_\mathrm{s}$, respectively.
The analytic solution is derived under the assumption that the ellipticity satisfies $|\epsilon|\ll1$, implying that the free-precession 
period is much longer than the spin period $P_\mathrm{n} = 2\pi/\Omega_\mathrm{n}$. 
To first order in $\epsilon$, the term $\pmb{\dot \omega}$ 
on the LHS of Eq. (\ref{eq:Euler_crust}) can therefore be neglected. 
Additionally, we consider that the external torque modifies the star's rotational evolution on a timescale much longer than the spin period
\begin{equation}
    |\pmb N_\mathrm{ext} |\ll I_0\Omega^2_\mathrm{n}.
\end{equation}
When the coupling of the superfluid and normal crust is sufficiently strong, the lag between the angular velocities of the superfluid and the normal crust,
$ \delta\pmb \Omega \equiv \pmb\Omega_\mathrm{s} - \pmb\Omega_\mathrm{n}$, is also small:
\begin{equation}
    |\delta\pmb \Omega| =|\pmb\Omega_\mathrm{s} - \pmb\Omega_\mathrm{n}| \ll \Omega_\mathrm{n}.
\end{equation}
At the same time, if coupling is weak or moderate, then $\delta\pmb\Omega$ will quickly drop to a small value, as shown below in Sect.~\ref{sec:analytic:steady}.
To leading order, the dissipative component of the internal torque can be approximated as:
\begin{equation}
    \pmb N_\mathrm{s1} \approx I_\mathrm{s}\Omega_\mathrm{n}\dfrac{{\cal R}}{1 + {\cal R}^2} \left [\delta\pmb \Omega + \left(\dfrac{\Omega_\mathrm{s}}{\Omega_\mathrm{n}}-1 \right)\pmb \Omega_\mathrm{n} \right].
\end{equation}
Under the above assumptions, and subtracting Eq. (\ref{eq:Euler_crust}) from (\ref{eq:Euler_superfluid}), 
one obtains the evolution equation for $\delta\pmb \Omega$:
\begin{equation}
\begin{split}
    \dfrac{\diff{(\delta\pmb \Omega)}}{\diff t} & = -(\pmb\omega \times\pmb\Omega_\mathrm{n}) -\dfrac{\pmb N_\mathrm{ext}}{I_0} \\
    & \quad - (\pmb\Omega_\mathrm{Sh} \times \delta\pmb \Omega) - \dfrac{\delta\pmb \Omega +(\Omega_\mathrm{s}/\Omega_\mathrm{n}-1)\pmb \Omega_\mathrm{n}}{\tau_\mathrm{d}},
\end{split}
    \label{eq:Delta_evolution}
\end{equation}
where 
\begin{equation}
    \tau_\mathrm{d} = \dfrac{1 + {\cal R}^2}{\Omega_\mathrm{n}{\cal R}(1 + \kappa_\mathrm{s})}
\end{equation}
is the relaxation time due to the dissipative internal torque component $\pmb N_{s1}$,  and 
\begin{equation}
    \pmb\Omega_\mathrm{Sh} = -\pmb\Omega_\mathrm{n}\,\dfrac{1- \kappa_\mathrm{s}{\cal R}^2}{1+{\cal R}^2}.
\end{equation}
\begin{figure*}
    \centering
    \includegraphics[width=\textwidth]{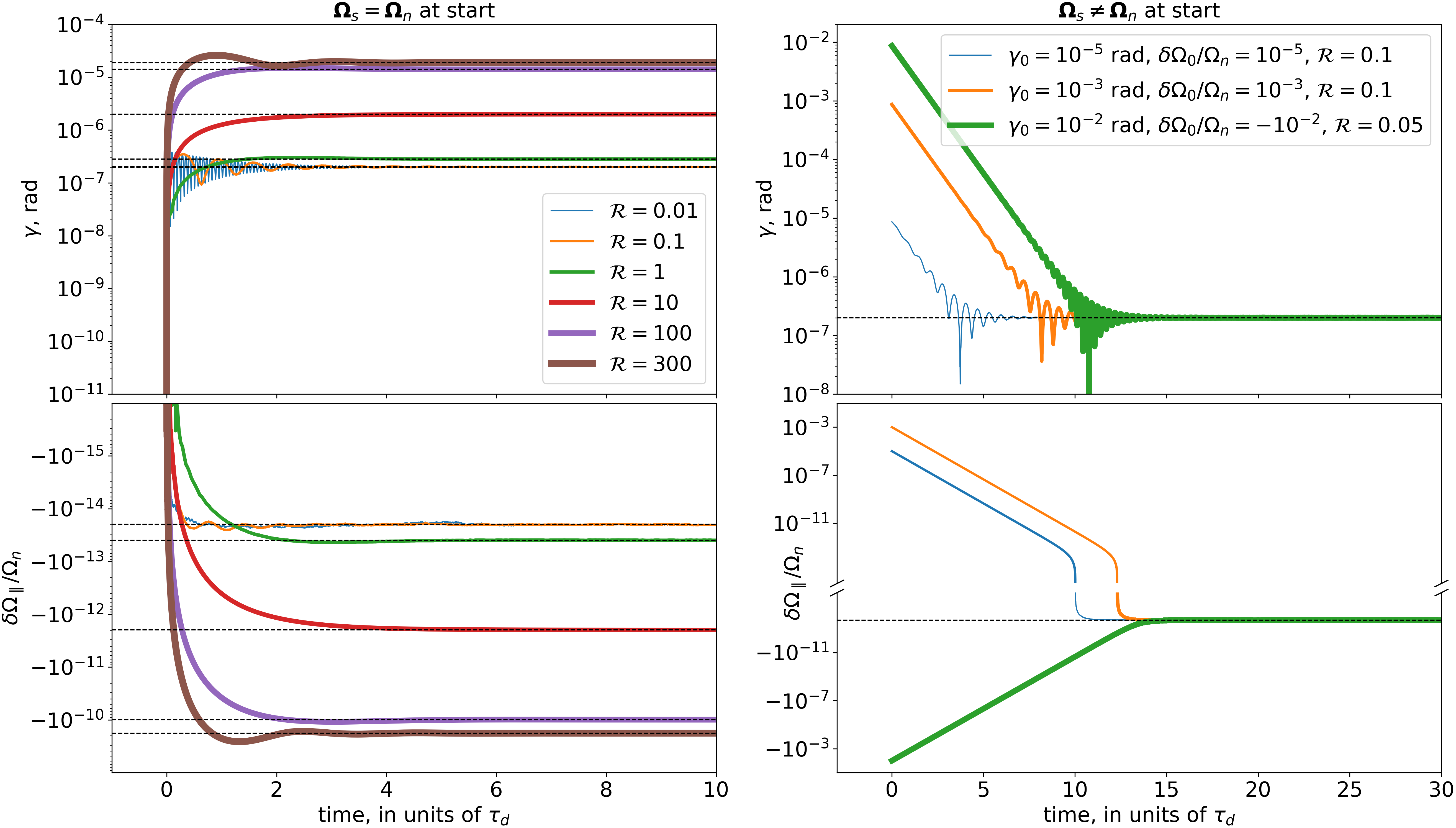}
    \caption{ Evolution of $\gamma$ and $\delta\Omega_{||}/\Omega_\mathrm{n} = \Omega_\mathrm{s}/\Omega_\mathrm{n}-1$ for free precession ($\pmb N_\mathrm{ext} = 0$), with $P_{\mathrm{n},0} = 1 \,\mathrm{s}$, $\epsilon = 4 \times 10^{-7}$, $\kappa_\mathrm{s}=0.01$, and initial tilt $\theta_0 = 45^\circ$.
    The left panels show solutions for different values of the drag parameter $\mathcal{R}$, assuming the initial condition $\pmb\Omega_\mathrm{n} = \pmb\Omega_\mathrm{s}$. The right panels show solutions for different initial values of $\gamma$ and $\delta\Omega_{||} =\Omega_\mathrm{s}-\Omega_\mathrm{n}$, as indicated.
    Solid lines represent numerical solutions of Eqs.~(\ref{eq:Euler_crust})–(\ref{eq:Euler_superfluid}), while horizontal dashed lines denote the analytic equilibrium values given by Eqs.~(\ref{eq:tilt_eq}) and (\ref{eq:lag_eq_2}).}
    \label{fig:steady_Rx}
\end{figure*}
Hereafter $\kappa_\mathrm{s} = I_0/I_\mathrm{s}$. Note that in the limit $\mathcal{R}\to \infty$, $\tau_\mathrm{d}\to \infty$ and $\Omega_\mathrm{Sh}$ reduces to wobble frequency derived by \cite{Shaham77}.
In the limit of zero drag, $\mathcal{R}\to 0$, the superfluid and crust are dynamically decoupled. The apparent rapid precession of the lag seen 
in the body frame is purely kinematic and results from the transformation to the rotating frame.  In the axi-symmetric case, $\pmb\Omega_\mathrm{s} || \pmb \Omega_\mathrm{n}$,
Eq.~(\ref{eq:Delta_evolution})reduces to $\dot{\delta\Omega} = N_\Omega/I_0 +2\delta\Omega/\tau_\mathrm{d}$, where $N_\Omega$ is the spin-down torque, and it is seen that in this case $\tau_\mathrm{d}/2$ is the damping time of the lag.

\subsection{Quasi-steady rotational evolution}
\label{sec:analytic:steady}

For a relaxation time much shorter than the spin-down time, $\tau_\mathrm{d}\ll (n_\Omega \Omega_\mathrm{n})^{-1}$,
the system relaxes to a quasi-steady state on a timescale $\sim \tau_\mathrm{d}$, so that for $t\gg\tau_\mathrm{d}$
Eq. \eqref{eq:Delta_evolution} satisfies $\dot{\delta\pmb\Omega}\approx0$ (see Appendix \ref{app:steady} for details). 
A convenient decomposition of the lag is
\begin{equation}
   \delta \pmb \Omega = \delta\pmb\Omega_\perp + \delta \pmb{\Omega}_{||},
\end{equation}
where
\begin{equation}
\begin{split}
   \delta \pmb\Omega_\perp &= \Omega_\mathrm{n}\, (\varphi_\mathrm{s} - \varphi)\sin\theta\cdot\pmb{e}_\varphi + \Omega_\mathrm{n}\,(\theta_\mathrm{s} - \theta)\cdot{\pmb e}_\theta, \\
    \delta\pmb{\Omega}_{||} &= (\Omega_\mathrm{s} - \Omega_\mathrm{n})\pmb{e}_\Omega,
\end{split}
\end{equation}
with the orthonormal basis vectors defined as:
$\pmb e_\Omega = \pmb\Omega_\mathrm{n}/\Omega_\mathrm{n}$,  $\pmb e_\theta = (\cos\theta\cos\varphi,\cos\theta\sin\varphi,-\sin\theta)^T$ and $\pmb e_\varphi = (-\sin\varphi,\cos\varphi,0)^T$.
To linear order $ |\delta\pmb\Omega_\perp|/\Omega_\mathrm{n}$ is the angle $\gamma$ between $\pmb \Omega_\mathrm{n}$ and $\pmb\Omega_\mathrm{s}$.
Solving Eq.~(\ref{eq:Delta_evolution}) for $\dot{\delta\pmb\Omega}=0$ gives the equilibrium components of $\delta\pmb\Omega$:
\begin{equation}
\begin{split}
& |\delta\pmb\Omega_\mathrm{\perp eq}|^2 \approx    \Omega_\mathrm{n}^2 \gamma^2_\mathrm{eq} =\\ 
& = \Omega_\mathrm{n}^2\left[(\epsilon\sin\theta\cos\theta + n_\varphi)^2 + n_\theta^2 \right]\dfrac{1 + {\cal R}^2}{1 + \kappa_\mathrm{s}^2 {\cal R}^2},
\end{split}
    \label{eq:tilt_eq}
\end{equation}
and
\begin{equation}
    \delta\Omega_{||\mathrm{eq}} = - \Omega_\mathrm{n}\dfrac{n_\Omega(1 + {\cal R}^2)}{2{\cal R}(1 + \kappa_\mathrm{s})}.
    \label{eq:lag_eq}
\end{equation}
Here $n_i$ are the dimensionless projections of the external torque, 
\begin{equation}
    n_{i}=\pmb N_\mathrm{ext}\cdot \pmb e_{i}/I_0\Omega_\mathrm{n}^2, \quad i=\Omega,\theta,\varphi.
    \label{eq:dimless_torques}
\end{equation}
Note that since $\epsilon, n_i \ll \kappa_\mathrm{s} \ll 1$ for a realistic neutron star, $\gamma_\mathrm{eq} \ll 1$ for any value of 
the drag parameter $\mathcal{R}$. The condition $\delta \Omega_\mathrm{||eq} \ll \Omega_\mathrm{n}$ is satisfied provided that 
$\Omega_\mathrm{n} \tau_\mathrm{d} \ll n_\Omega^{-1} $. In case of free precession, $n_\Omega=0$, the equilibrium value of  
$\delta\Omega_\mathrm{||}$ must be determined to second order. Omitting the details, one finds that
\begin{equation}
    \delta\Omega_{||\mathrm{eq}} = - \Omega_\mathrm{n}\dfrac{n_\Omega(1 + {\cal R}^2)}{2{\cal R}(1 + \kappa_\mathrm{s})} - \Omega_\mathrm{n}\dfrac{\gamma^2_\mathrm{eq}}{2(1+\kappa_\mathrm{s})}.
    \label{eq:lag_eq_2}
\end{equation}
For realistic stars $n_\Omega \gg \gamma_\mathrm{eq}^2$, so that a linear expansion provides a good approximation for the lag. The second-order term is included here only to enable comparison between the analytical and numerical solutions in the case of free precession.

The relaxation toward quasi-steady evolution in the case of free precession is illustrated in Fig.~\ref{fig:steady_Rx}, which shows numerical solutions of Eqs.~(\ref{eq:Euler_crust}) and (\ref{eq:Euler_superfluid}) for different parameter values.  It indicates that the characteristic time to reach the 
quasi-steady state is  $\tau_\mathrm{d}$, as
naively expected. A good agreement is seen between the analytic values of $\gamma_\mathrm{eq}$
and $\delta \Omega_\mathrm{eq}$ given in Eqs.~\eqref{eq:tilt_eq} and \eqref{eq:lag_eq_2} (the horizontal dashed lines in Fig.~\ref{fig:steady_Rx}), and the numerical solutions. 

\begin{figure}
    \centering
    \includegraphics[width=\columnwidth]{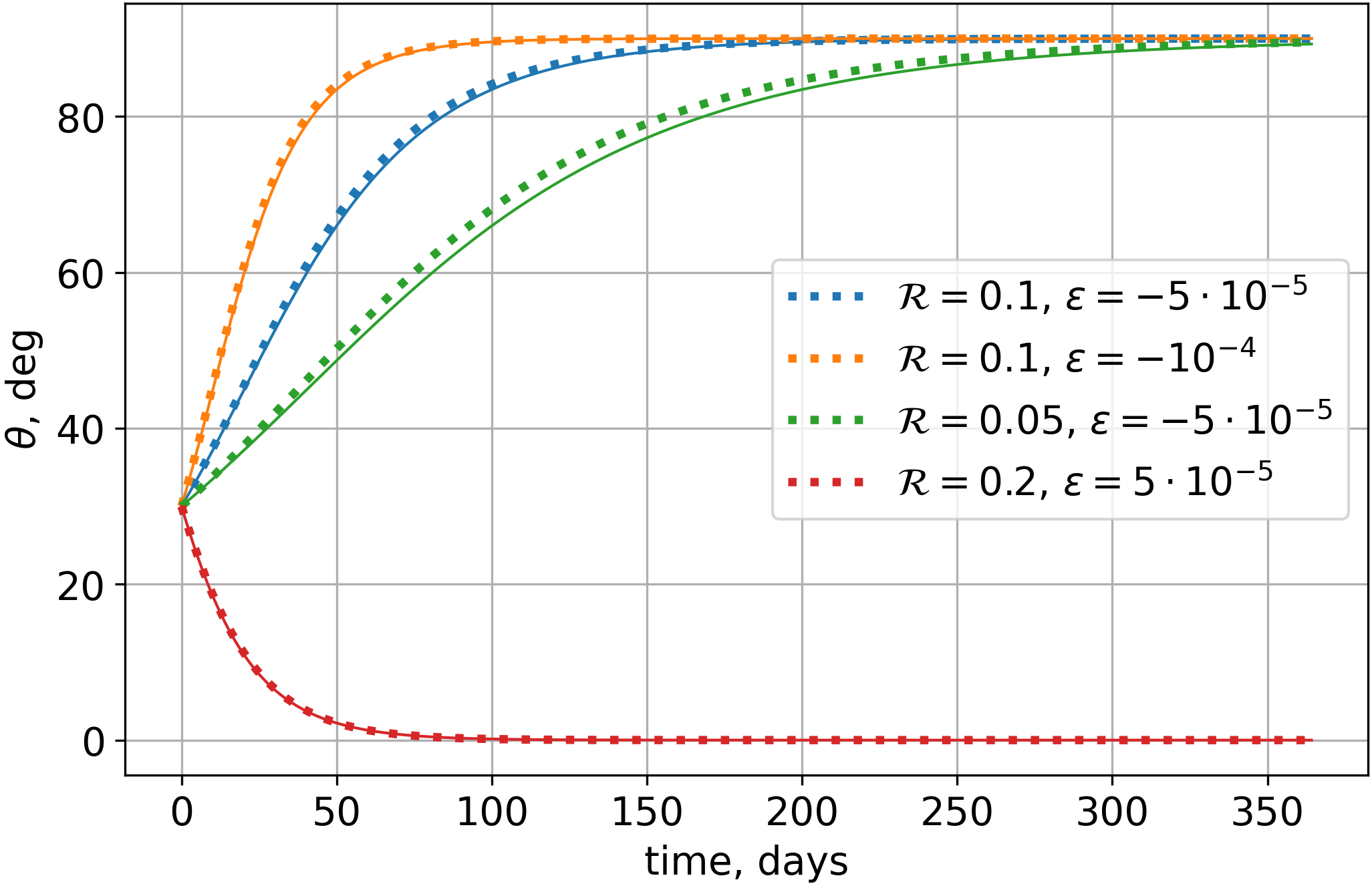}
    \caption{The evolution of the wobbling angle $\theta$ in the limit of free precession, for $\kappa_s=10^{-2}$, and for different values of the drag parameter $\mathcal{R}$ and ellipticity $\epsilon$, as indicated. The initial crustal spin period and tilt are $P_{n,0}=1 \;\mathrm{s}$ and $\theta_0=30^\circ$, respectively. The dotted lines show the exact numerical solutions of Eqs.~(\ref{eq:Euler_crust}) and (\ref{eq:Euler_superfluid}), while the solid lines show the corresponding analytical solutions given by Eq.~(\ref{eq:free_damp_theory}). Good agreement is obtained.}
    \label{fig:free_damp}
\end{figure}

Upon substituting $\pmb\Omega_\mathrm{s} = \pmb\Omega_\mathrm{n} + \delta\pmb\Omega_\mathrm{eq}$ into Eq.~(\ref{eq:Euler_crust}), one obtains the secular evolution equations for the wobbling angle $\theta$ and the azimuthal angle $\varphi$:
\begin{align}
 \dot\theta &= \Omega_\mathrm{n}\dfrac{n_\theta - (\epsilon\sin\theta\cos\theta + n_\varphi)\kappa_\mathrm{s}{\cal R}}{1 + \kappa_\mathrm{s}^2{\cal R}^2}, 
    \label{eq:dot_theta}   \\
  \dot\varphi &= \Omega_\mathrm{n}\dfrac{\epsilon\sin\theta\cos\theta + n_\varphi + n_\theta\kappa_\mathrm{s}{\cal R}}{\sin\theta(1 + \kappa_\mathrm{s}^2{\cal R}^2)}, \label{eq:dot_phi}  
\end{align}
restricted to the range $ \mathcal{R} \ll (1+\kappa_s)/n_\Omega$.
As will be clarified below, Eq. (\ref{eq:dot_theta}) governs the decay of slow precession modes, owing to the coupling of the crustal 
superfluid and the normal crust.  The decay of the fast (Shaham) mode occurs during the rapid transient phase, over time 
$t\lesssim \tau_\mathrm{d}$ (see Appendix \ref{app:steady}). The spin-down equation for the normal crust (and the coupled core) is obtained, to linear order, by 
projecting Eq.~(\ref{eq:Euler_crust}) onto $\pmb e_\Omega$:
\begin{equation}
\begin{split}
    \dot{\Omega}_\mathrm{n} & = \Omega^2_\mathrm{n}\, n_\Omega + \dfrac{2\kappa_\mathrm{s}{\cal R}}{1 + {\cal R}^2} \Omega_\mathrm{n} \delta\Omega_\mathrm{||eq} \\
    & = \dfrac{\Omega^2_\mathrm{n}n_\Omega}{1+\kappa_\mathrm{s}} = \frac{\pmb N_\mathrm{ext}\cdot\pmb e_\Omega}{I_\mathrm{0}+I_\mathrm{s}}.
\end{split}
\end{equation}
We see that the spin-down rate of the crust coincides with that of a rigid body, as required by angular momentum conservation. 
%
%
It is instructive to first consider the limit of free precession, as it captures, to a good approximation, the evolution of the system in the general case when $n_i \ll \epsilon$, as in the the case of Her X-1. In this limit, Eqs.~\eqref{eq:dot_theta} and \eqref{eq:dot_phi} reduce to
\begin{equation}
    \dot\varphi  = \dfrac{\epsilon \Omega_\mathrm{n}\cos\theta}{1 + \kappa_\mathrm{s}^2{\cal R}^2},
\end{equation}
and 
\begin{equation}
    \dot\theta  
    = -\dfrac{\epsilon\Omega_\mathrm{n} \kappa_\mathrm{s}{\cal R}}{1 + \kappa_\mathrm{s}^2{\cal R}^2}\, \cos\theta\sin\theta.
    \label{eq:free_precession_damp}
\end{equation}
The solution of the last equation can be written as:
\begin{equation}
    \tan(\theta) = \tan(\theta_0) \cdot \exp\left(-\dfrac{t}{\tau_\mathrm{p}}\right),
    \label{eq:free_damp_theory}
\end{equation}
where
\begin{equation}
    \tau_\mathrm{p} = \dfrac{1 + \kappa_\mathrm{s}^2{\cal R}^2}{\Omega_\mathrm{n}\epsilon\kappa_\mathrm{s}{\cal R}}
    \approx \frac{(1+\kappa^2_s\mathcal{R}^2)}{\epsilon\kappa_s(1+\mathcal{R}^2)}\tau_\mathrm{d} > \frac{\kappa_s\tau_\mathrm{d}}{\epsilon}.
\end{equation}
As seen, for $\epsilon \ll \kappa_s$ we have $\tau_\mathrm{p} \gg \tau_\mathrm{d}$.
Note that the evolution of $\theta$ depends on the sign of $\epsilon$ (see Fig.~\ref{fig:free_damp}). For $\epsilon>0$ (an oblate star), $\theta \to 0$ as $t \gg \tau_\mathrm{p}$, and the amplitude of the precession decays. For $\epsilon<0$ (a prolate star), $\theta \to \pi/2$ as $t \gg \tau_\mathrm{p}$, while $\dot\varphi \to 0$, i.e. the precession period diverges. The evolution of $\theta$  reflects the tendency of the system to minimize its rotational energy at fixed angular momentum.
In either case, the slow free–precession mode damps on a timescale $\sim \tau_\mathrm{p}$. In other words, the slow precession persists for $\gtrsim (\kappa_\mathrm{s}{\cal R})^{-1}$ cycles independently on $\epsilon$. 

\section{Application to Hercules X-1}
\label{sec:HerX1}

The origin of the $\sim35$-day modulation observed in the X-ray binary Her X-1 \citep{Tananbaum1972, Giacconi1973} has been debated for several decades. A widely accepted interpretation attributes the cycle to the forced precession of a tilted and warped accretion disk, which periodically obscures the central X-ray source and produces the characteristic sequence of main-on, short-on and off states in the X-ray light curve \citep{katz1973,petterson1975,SchandlMeyer1994,wijers1999}. In this picture, tidal torques from the companion star together with radiation-driven warping lead to a precessing disk whose varying orientation modulates the observed flux. An alternative interpretation invokes free precession of the neutron star, motivated in part by the systematic evolution of the X-ray pulse profiles over the 35-day cycle, which may reflect a changing orientation of the emission beam relative to the observer \citep{Truemper1986,ShakuraPostnovProkhorov1998,Postnov2013,Kolesnikov2022}. Recent X-ray polarimetric observations with the Imaging X-ray Polarimetry Explorer have provided new evidence supporting this possibility. Phase-resolved polarization measurements across the superorbital cycle reveal systematic variations in the polarization angle and degree that are consistent with changes in the viewing geometry of the magnetic poles expected if the neutron star itself undergoes free precession \citep{Heyl24,Zhao2024}. These observations suggest that the $\sim35$-day clock may be set by the free precession of the neutron-star crust, implying a small but finite asymmetry in the star’s moment of inertia (of order a few parts per $10^{7}$) required to sustain such motion. Hybrid scenarios have also been proposed in which both the neutron star and the accretion disk precess, forming a system of coupled clocks in which the disk primarily controls the X-ray turn-ons while neutron-star precession modulates the pulse-profile morphology \citep[e.g.,][]{Postnov2000,Staubert2009}. The relative roles of these mechanisms remain an active topic of investigation.

In what follows, we consider the implications of neutron star precession for the superfluid dynamics.
Under the assumption of tight core-crust coupling, the net external torque acts on the star on timescales much longer than the observed 35-d cycle in Her X-1, which in the free-precession interpretation is driven by a stellar deformation with ellipticity $\epsilon\approx 4\times10^{-7}$ \citep[e.g.,][]{Heyl24}. More precisely,
\begin{equation}
    N_\mathrm{ext,HX1} \ll I_0\Omega_\mathrm{n}^2 |\epsilon|,
\end{equation}
where $N_\mathrm{ext,HX1}$ is the magnitude of the net external torque, which can be decomposed into several components,
\begin{equation}
  {\pmb N}_\mathrm{ext,HX1} = \pmb N_\mathrm{acc} + \pmb N_\mathrm{mag} + \pmb N_\mathrm{psr} + 
  \pmb N_\mathrm{warp} + \pmb N_\mathrm{prec}.
\end{equation}
From left to right, the terms on the right-hand side represent the accretion spin-up torque, magnetospheric spin-down torque, pulsar losses, and the back-reaction associated with the warping and precession of the accretion disc, respectively. The latter two are due to the interaction of the inclined disc with the NS magnetosphere \citep{2011MNRAS.412.2790L,Lai99}.
The details are provided in Appendix~\ref{app:HerX1_torque}.

It is worth noting that ${\pmb N}_\mathrm{acc}$, ${\pmb N}_\mathrm{warp}$, and ${\pmb N}_\mathrm{prec}$ rotate rapidly about $\pmb \Omega_\mathrm{n}$ in the principal frame of the star, as they are linked to the accretion disc. Thus, their contribution effectively averages out over the spin period. The corresponding spin-averaged torque can be written as the sum of two components, parallel and orthogonal to the crust's spin axis, respectively:
\begin{equation}
    \langle \pmb N_\mathrm{ext,HZ1}\rangle_\mathrm{spin} = \pmb N_\parallel + \pmb N_\perp.
    \label{eq:HerX1_averaged_torque}
\end{equation}
The first component is responsible for the neutron star spin-down:
\begin{equation}
\begin{split}
    {\pmb N}_\parallel & = \Biggl [ N_\mathrm{psr,0}(1 + \sin^2\chi) + N_\mathrm{mag} \Biggr . \\ 
    & \Biggl . + N_\mathrm{acc,0}\cos\alpha \left(1 + \dfrac{4\sqrt{2}}{3}\sin^2\alpha\cos^2\chi \right) \Biggr] \pmb e_\Omega,
\end{split}
    \label{eq:NHZ_parallel}
\end{equation}
while the second one is capable of changing the orientation of $\pmb\Omega_\mathrm{n}$ with respect the principal frame:
\begin{equation}
    \pmb N_\perp = N_\chi \left ((\pmb m \times \pmb e_\Omega) \times \pmb e_\Omega \right) + N_\mathrm{rad}(\pmb m \times \pmb e_\Omega),
    \label{eq:NHZ_perp}
\end{equation}
Here $\pmb m$ is the unit vector aligned with the magnetic axis. It is assumed to be fixed in the principal frame, and misaligned with respect to the deformation axis. The torque
\begin{equation}
    N_\chi = \left(N_\mathrm{psr,0} + \eta\,A\,  N_\mathrm{acc,0}\sin^2\alpha\cos\alpha \right)\cos\chi
    \label{eq:NHZ_perp_chi}
\end{equation}
is responsible for the star magnetic obliquity evolution, while 
\begin{equation}
    N_\mathrm{rad} = \dfrac{N_\mathrm{psr,0}}{3} \left( \dfrac{r_\mathrm{LC}}{r_\mathrm{NS}} \right) \cos\chi
\end{equation}
is the pulsar ``anomalous'' torque \citep[e.g][]{bz_anomal2014}.
%
Here $\chi = \cos^{-1}(\pmb m\cdot \pmb e_\Omega)$ is the magnetic angle, $\alpha$ is the angle between the accretion-disk axis and $\pmb\Omega_\mathrm{n}$, $r_\mathrm{LC} = c/\Omega_\mathrm{n}$ is the light-cylinder radius, and $r_\mathrm{NS}$ is the neutron-star radius. The dimensionless parameters $\eta = 0..1$ and 
\begin{equation}
    A = \left[1- \dfrac{\eta}{2}\left(\sin^2\chi\sin^2\alpha + 2\cos^2\chi\cos^2\alpha \right) \right]^{-1},
\end{equation}
describe the spin-modulated accretion torque.

The orthogonal component, $\pmb N_\perp$, is the only one capable of influencing the time evolution of the wobbling angle, $\dot\theta$. Indeed 
\begin{equation}
    n_{\theta,\varphi} \equiv \dfrac{N_{\theta,\varphi}}{I_0\Omega_\mathrm{n}^2} = \dfrac{\pmb N_\perp\,\cdot\pmb e_{\theta,\varphi}}{I_0 \Omega_\mathrm{n}^2},
\end{equation}
while $\pmb N_\perp\cdot \pmb e_\Omega \equiv 0$, and $n_\Omega \equiv N_\parallel/I_0\Omega_\mathrm{n}^2$. 
For the estimates below, we adopt parameters consistent with those inferred for the neutron star in Her~X-1. In particular, we take a 
spin period $P_\mathrm{n,HX1} = 1.2$~s, a magnetic moment $\mu_\mathrm{HX1} = 3 \times 10^{30}$~G~cm$^3$ \citep{Staubert2014}, and a 
magnetospheric radius $r_\mathrm{m} = 10^3$~km. The latter is a conservative order-of-magnitude estimate that remains smaller than the 
corotation radius of the system, $r_\mathrm{c} \approx 2 \times 10^3$~km (for a $1.5\,M_\odot$ neutron star). Since $r_\mathrm{m} < 
r_\mathrm{c}$, the system is expected to remain in the accretion regime. However, $r_\mathrm{m}$ for Her X-1 is not well constrained 
observationally; estimates in the literature span a broad range, from $\sim 100$ to $4 \times 10^3$~km \citep{Averintsev1992, Scott2000, 
Ramsay2002}. Within the adopted torque model (Appendix~\ref{app:HerX1_torque}), $r_\mathrm{m} = 10^3$ km corresponds to an accretion 
rate $\dot M_\mathrm{HX1} = 5\cdot 10^{-8}$ $M_\odot$ yr$^{-1}$.
\begin{figure*}
    \centering
    \includegraphics[width=\textwidth]{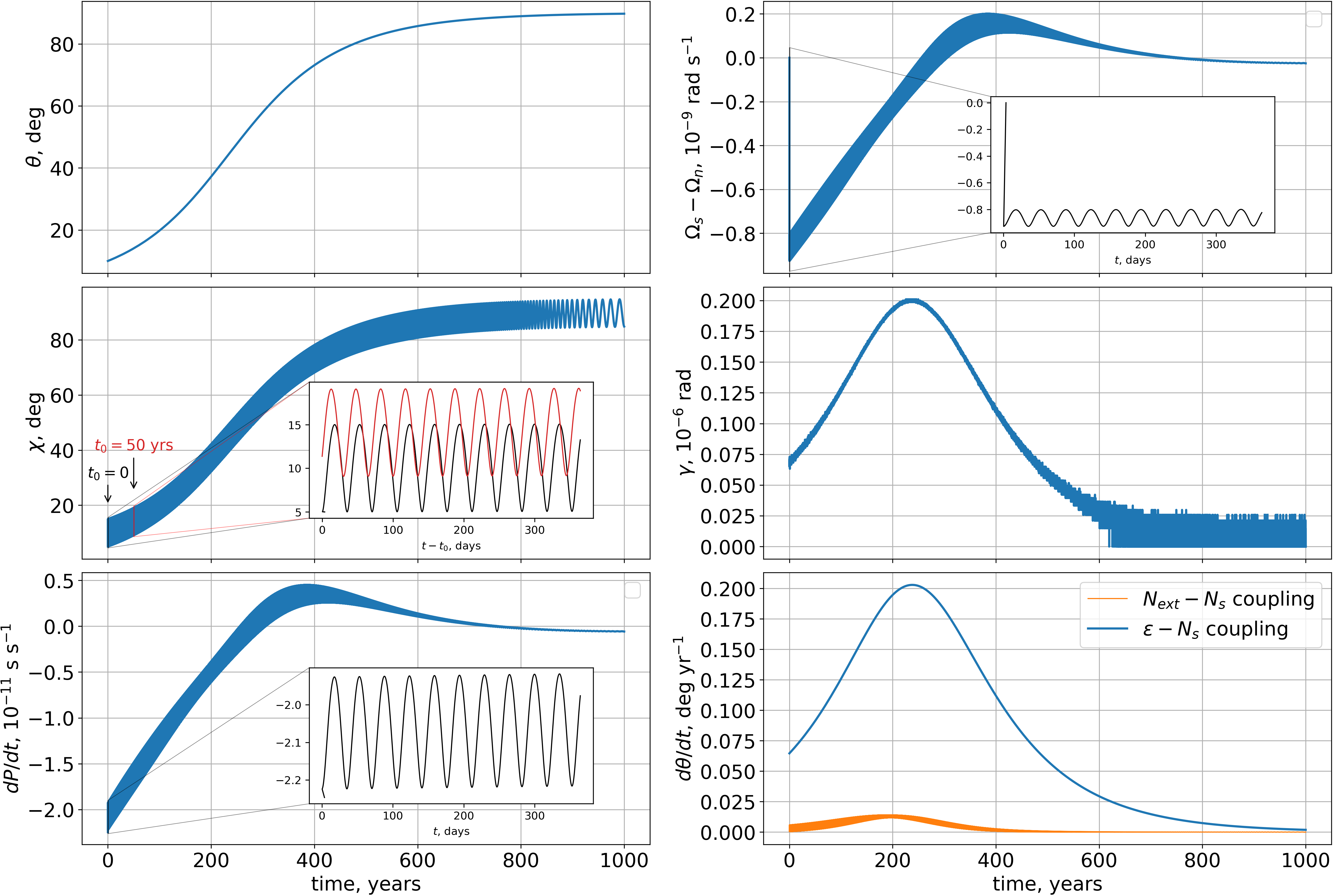}
    \caption{Possible evolutionary track of the neutron star in Her X-1, taking into account the internal torque exerted by the crustal superfluid. The plots show the numerical solution of Eqs.~(\ref{eq:Euler_crust})–(\ref{eq:Euler_superfluid}) including the external torque given by Eq.~(\ref{eq:HerX1_averaged_torque}). The star is assumed to be prolate with ellipticity $\epsilon = -4\times10^{-7}$ and superfluid drag coefficient ${\cal R}=\kappa_\mathrm{s}=0.01$; the accretion rate is $\dot M = 5\times10^{-8}\,M_\odot\,\mathrm{yr}^{-1}$.
In the left column, the evolution of the wobbling angle $\theta$, magnetic angle $\chi$, and spin-down rate $\dot{P}$ are shown (from top to bottom). In the right column, the angular-frequency lag $\delta\Omega_{||}=\Omega_\mathrm{s}-\Omega_\mathrm{n}$ and the tilt $\gamma$ are shown in the top and middle panels, respectively. The bottom-right panel shows the contributions to $\dot\theta$ described by Eq.~(\ref{eq:dot_theta}): the term $\propto -\epsilon\sin\theta\cos\theta\,\kappa_\mathrm{s}{\cal R}$ is shown by the blue curve, while the term $\propto n_\theta - n_\varphi\kappa_\mathrm{s}{\cal R}$ is shown by the orange curve. The evolution of the wobbling angle is therefore predominantly driven by the superfluid torque coupled to the stellar deformation.}
    \label{fig:HerX1_R001}
\end{figure*}
\begin{figure*}
    \centering
    \includegraphics[width=\textwidth]{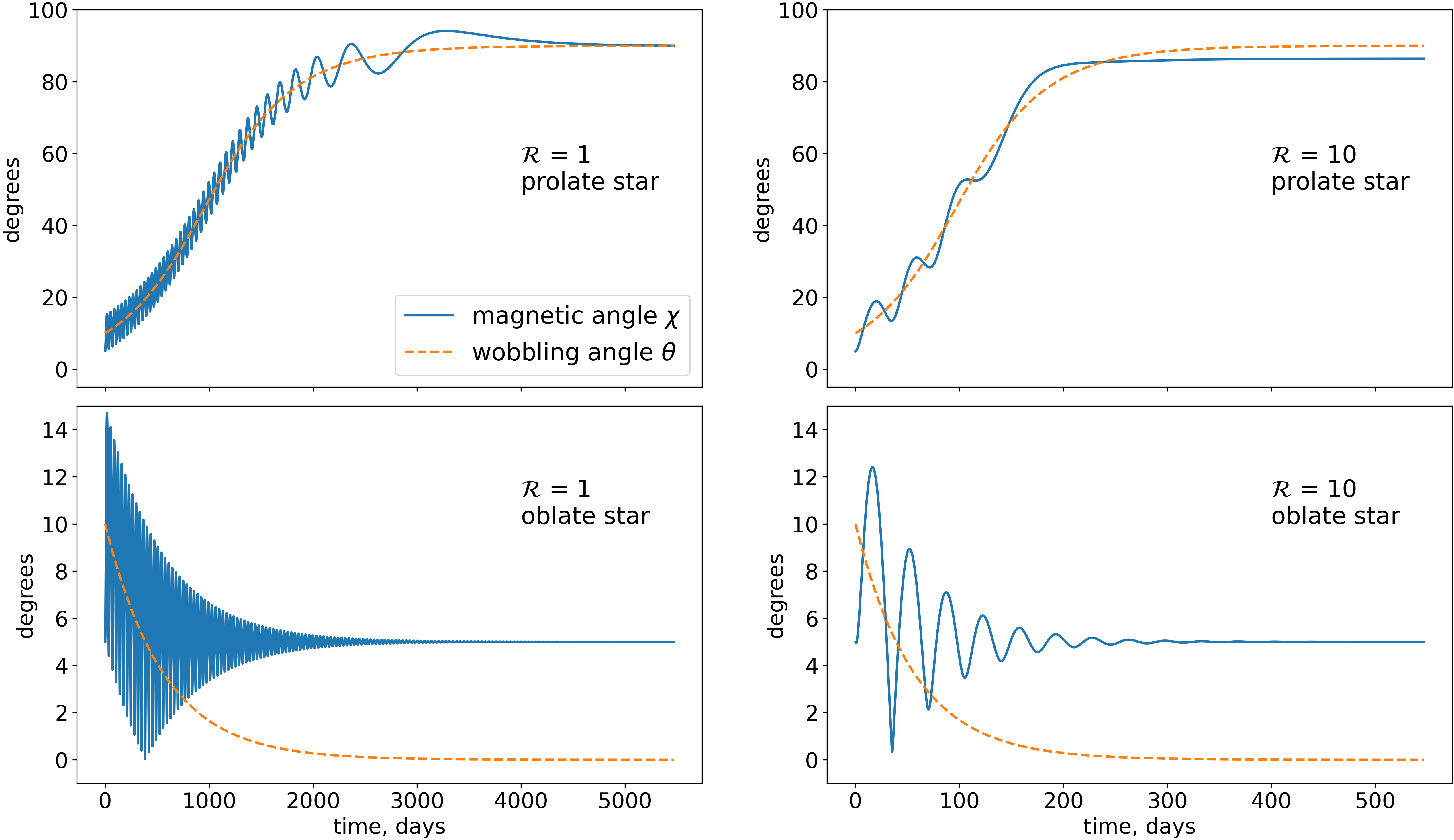}
    \caption{Modeled evolution of the magnetic angle $\chi$ \textbf{and the wobbling angle $\theta$} of Her X-1 for ${\cal R}=1$ (left panels) and ${\cal R}=10$ (right panels), for prolate (top) and oblate (bottom) stars. All the other parameters are the same as in Figure~\ref{fig:HerX1_R001}.}
    \label{fig:HerX1_Rcompare}
\end{figure*}
This value is about an order of magnitude larger than observational estimates \citep{Boroson2007,Kosec2020}; however, since the
spin-down timescale of the star is much longer than the precession period, the resulting torque estimates remain representative for our 
purposes. Namely,
\begin{equation}
    N_\mathrm{acc,0} = (4.7\times 10^{35}\mbox{ dyn cm})\, \left(\dfrac{\mu}{\mu_\mathrm{HX1}} \right)^{2/7}\, 
    \left(\dfrac{\dot M}{\dot M_\mathrm{HX1}}\right)^{6/7},
\end{equation}
%
%
\begin{equation}
    N_\mathrm{mag} = (-4.1 \times 10^{35}\mbox{ dyn cm})\,\left(  \dfrac{\mu}{\mu_\mathrm{HX1}} \right)^2\,
    \left(\dfrac{P}{P_\mathrm{n,HX1}} \right)^{-2},
\end{equation}
%
and
\begin{equation}
    N_\mathrm{psr,0} = (-4.8\times 10^{31}\mbox{ dyn cm})\, \left(  \dfrac{\mu}{\mu_\mathrm{HX1}} \right)^2 \left(\dfrac{P}
    {P_\mathrm{n,HX1}} \right)^{-3}.
\end{equation}
%
Notice that $(r_\mathrm{LC}/r_\mathrm{NS})N_\mathrm{psr,0}/3 \approx -10^{35}$ dyn cm for $r_\mathrm{NS} = 10$ km, implying that the pulsar torque can contribute equally to the precessional evolution of the star.

For the order-of-magnitude estimates below, we neglect geometric factors and adopt
$N_\mathrm{ext,HX1}\sim N_\perp \sim N_\parallel \sim 10^{35}$ dyn cm, given that
$I_0 = 10^{45}$ g cm$^2$, $P_\mathrm{n} = 1.2$ s, and $|\epsilon| = 4\cdot10^{-7}$ \citep{Heyl24}. Thus one gets
\begin{equation}
    \dfrac{N_\mathrm{ext,HX1}}{I_0\Omega_\mathrm{n}^2 |\epsilon|} \approx 10^{-5}, 
\end{equation}
so that the external torque is negligible compared to the deformation torque.  
On the other hand, as total $\dot\theta \propto n_\theta - (\epsilon\sin\theta\cos\theta + n_\varphi)\kappa_\mathrm{s}{\cal R}$, then for
\begin{equation}\label{E:R}
    {\cal R} \gtrsim 0.002\,N_{\perp,35}\,|\epsilon|_{-7}^{-1}\,\kappa^{-1}_\mathrm{s,-2}\,I_{0,45}^{-1}\,P_\mathrm{n,s}^{2},
\end{equation}
the internal torque is also substantially stronger than the external one. Here, $N_{\perp,35} = N_{\perp}/(10^{35}\mbox{ dyn cm})$, $|\epsilon|_{-7}=|\epsilon/10^{-7}|$, $I_{0,45} = I_0/(10^{45}\mbox{ g cm}^2)$ and $\kappa_\mathrm{s,-2} = \kappa_\mathrm{s}/10^{-2}$.

In the case of weak pinning (Eq.~\ref{E:R} is not fulfilled), and if $\epsilon\cos\theta$ and $n_\theta$ have the same signs, the latter can be large enough to compensate, in principle, the dissipation of 
free precession due to crust-superfluid interaction.

This might be possible as $\pmb N_{\rm ext}$ consists of components 
which affect the orientation of the rotation axis 
with respect to the principal axes of the crust. One is the last term in $\pmb N_\perp$ related to pulsar emission, and the other is the term proportional to $\eta$ which arises from the accretion torque modulation with the spin phase (see Eq.~\ref{eq:NHZ_perp_chi}). 

However, a detailed analysis of the specific conditions that must be met for such a balance is beyond the scope of this work.

In the opposite case, if external torque does not compensate the dissipation, the neutron star precesses essentially freely, and the 
evolution of $\theta$ is controlled by the superfluid component. Even for ${\cal R} = 0.01$, the precession of Her X-1 neutron star 
damps on a timescale of at least $(\kappa_\mathrm{s}{\cal R})^{-1} \approx 10^4$ cycles, corresponding to $\approx 10^3$ years. This is much longer than the $\sim50$-year observational time span of the system.

To demonstrate this, we solved Eqs.~(\ref{eq:Euler_crust}) and (\ref{eq:Euler_superfluid}) including the external torque 
(\ref{eq:HerX1_averaged_torque}), adopting $\chi_0 = 5^\circ$ (with the magnetic axis tilted relative to the deformation axis by an angle $\theta_m = 10^\circ$) and $\alpha_0 = 25^\circ$. These values are consistent with the parameters provided for the Her X-1 system \citep[see e.g.][]{Heyl24}. The amplitude of the magnetic angle variations during the 35-day cycle, estimated from X-ray polarization, is only a few degrees, implying that the crust spin axis and the magnetic axis are nearly aligned with the symmetry axis. We also assumed a prolate star ($\epsilon = -4\times10^{-7}$) with small loss of generality. On the other hand, negative $\epsilon$ could be justified while the internal toroidal magnetic field tends to distort a star into a prolate shape \citep{cutler2002}.

The spin–disc tilt $\alpha$ was evolved self-consistently as
\begin{equation}
I_0 \Omega_\mathrm{n}\,\dot{\alpha}
= -N_{\mathrm{acc},0}\sin\alpha
\left(1 - \frac{4\sqrt{2}}{3}\cos^2\alpha\cos^2\chi \right).
\end{equation}
The superfluid parameters were set to ${\cal R}=\kappa_\mathrm{s}=0.01$, and the accretion torque modulation parameter to $\eta=0.99$.
The system has evolved over 1000 years.

Figure~\ref{fig:HerX1_R001} shows the results. The left column displays $\theta$, $\chi$, and $\dot{P}$ (top to bottom), while the right 
column shows the crust–superfluid lag $\delta\Omega$ and tilt $\gamma$. The bottom-right panel decomposes $\dot\theta$ into the internal 
contribution $\propto -\epsilon \sin\theta\cos\theta\,\kappa_\mathrm{s}{\cal R}$ (blue) and the external torque term $\propto n_\theta - 
n_\varphi \kappa_\mathrm{s}{\cal R}$ (orange). For these parameters, the external torque has a negligible impact on $\theta$, and the 
system follows a free-precession-like evolution. Since $\epsilon<0$, the wobbling angle grows secularly on $\sim 10^2$-yr timescales, 
consistent with Eq.~(\ref{eq:dot_theta}), while $\chi$ exhibits nearly constant modulation amplitude with a slow drift in its cycle-averaged value,
$\langle \chi \rangle$, by $\sim 5^\circ$ over 50 years. Spin-down modulations (Fig.~\ref{fig:HerX1_R001}, lower left) are dominated by 
the warping torque term in Eq.~(\ref{eq:NHZ_parallel}) and exceed observational estimates by roughly an order of magnitude
(cf. Fermi GBM data\footnote{\texttt{https://gammaray.nsstc.nasa.gov/gbm/science/pulsars/lightcurves/herx1.html}}; 
\citealt{GBM2009,Kolesnikov2022}), reflecting that the parameters are not fully tuned to observations.
Nevertheless, even this enhanced accretion torque does not alter the precessional evolution, since the internal contribution to 
$\dot\theta$ remains dominant.

For completeness, we also computed the rotational evolution of Her X-1 using the same parameters but with stronger superfluid drag, \({\cal R}=1\) and \(10\). The results are shown in the left and right columns of Fig.~\ref{fig:HerX1_Rcompare}, respectively. We further considered both prolate (\(\varepsilon=-4\times10^{-7}\)) and oblate (\(\varepsilon=+4\times10^{-7}\)) configurations to illustrate the effect of precession damping predicted by Eq.~(\ref{eq:free_precession_damp}). 
The results are fully consistent with the analytical expectations: increasing drag accelerates the evolution of the precession, driving the wobbling angle toward either \(0^\circ\) or \(90^\circ\), depending on the sign of \(\epsilon\). 
\begin{figure*}
    \centering
    \includegraphics[width=\textwidth]{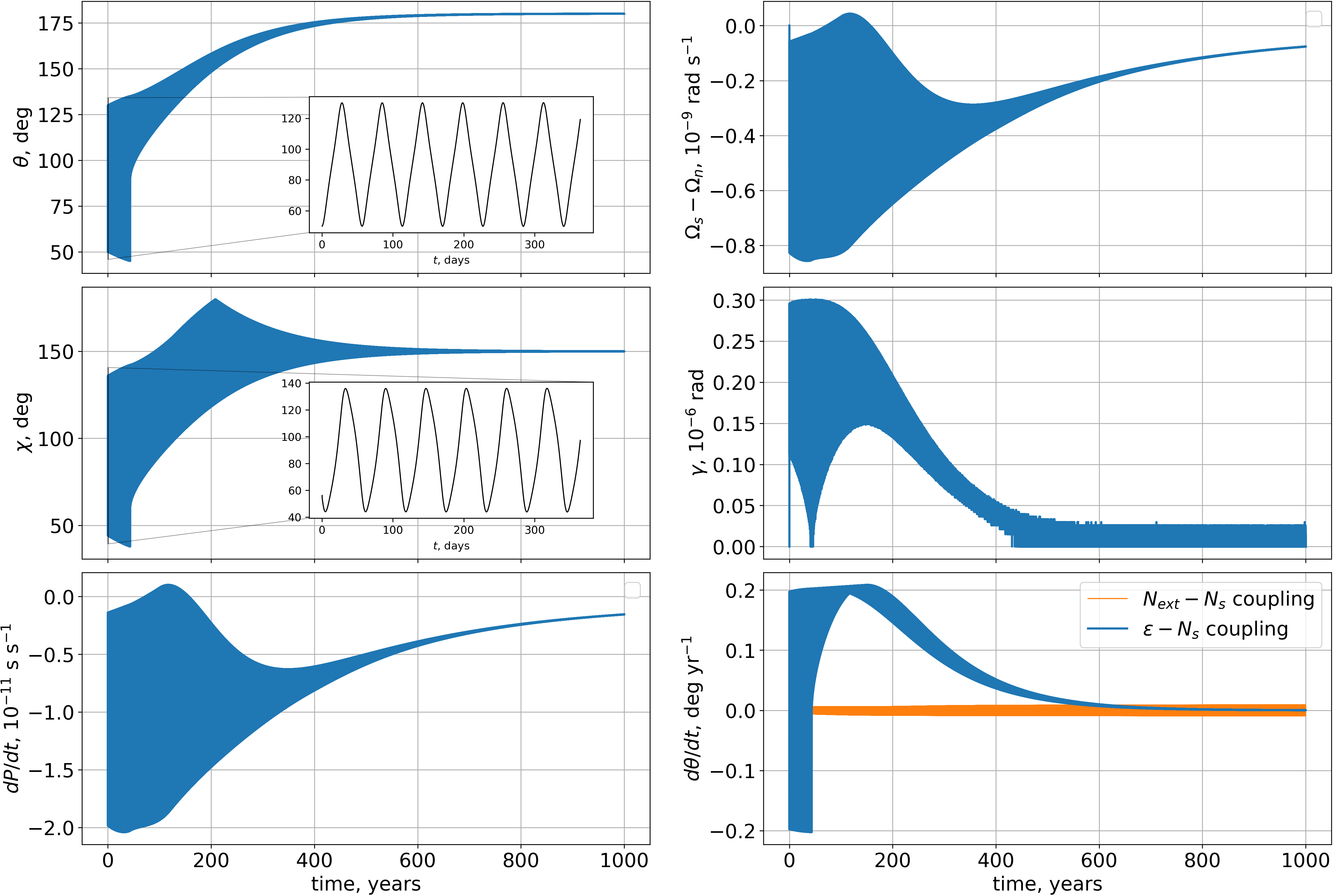}
    \caption{Possible evolution of the neutron star in Her X-1 assuming a triaxial deformation $\epsilon_3 = -\epsilon_1 = 4\times10^{-7}$ and initial orientations of $\pmb{\Omega}_\mathrm{n}$ and $\pmb{m}$ as estimated by \cite{Kolesnikov2022}. In particular, $\theta_0 = 50^\circ$, and $\pmb{m}$ points in the direction $(\varphi_\mathrm{m},\theta_\mathrm{m}) = (90^\circ, 30^\circ)$.  The terms contributing to $\dot{\theta}$ in the bottom right plot were calculated using the Eq.(\ref{eq:dot_theta}) assuming $\epsilon = \sqrt{\epsilon_1^2 + \epsilon_3^2}$.  Although this equation is formally derived for the biaxial case and therefore does not strictly apply to a triaxial star, it still provides a reasonable theoretical estimate of $\dot{\theta}$.  All other parameters are the same as in Figure~\ref{fig:HerX1_R001}. The stellar evolution is qualitatively similar to the biaxial case; during the first $\sim100$ years the spin axis precesses around the minor inertia axis ($\epsilon_1$), with large-amplitude variations of $\theta$ and $\chi$. Subsequently, it transitions to precession about the major inertia axis ($\theta \rightarrow 180^\circ$) on a timescale determined by the dissipative internal torque.}   
     \label{fig:HerX1_R001_triaxial}
\end{figure*}
In the calculations above, the Her X-1 neutron star was assumed to be biaxial, which is not necessarily the case. For example, \cite{Kolesnikov2022} argued that the observed 35-d spin variations are consistent with triaxial precession of the star. Analyzing pulsar spin-down history, they estimate the system parameters as \(\theta = 50^\circ\), with the magnetic axis oriented along \((\varphi_\mathrm{m},\theta_\mathrm{m}) = (90^\circ,30^\circ)\). Adopting these parameters as an initial rotational state, along with $\varepsilon_3 = -\varepsilon_1 = 4\times10^{-7}$, and keeping all other parameters the same as used for the biaxial case, we performed simulations of the spin evolution for a triaxial configuration. The results are shown in Fig.~\ref{fig:HerX1_R001_triaxial}.

The rotational fate of the star is qualitatively similar to that in the biaxial case. The spin axis aligns with one of the principal axes on the timescale set by the dissipative component of the internal torque. During the first \(\sim 100\) years of evolution, the star precesses around the minor inertial axis with large-amplitude variations in \(\theta\) and \(\chi\). It then transitions to a different precessional state on an orbit around the principal axis with the maximal moment of inertia. This behavior is fully consistent with expectations for triaxially deformed neutron stars with dissipative torques; see \cite{Biryukov2025} for a detailed theoretical discussion. 

Our final conclusion is that for sufficiently weak drag, \({\cal R} \lesssim 0.01\), the neutron star in Her X-1 can sustain precession for at least hundreds of years, consistent with observational constraints.   Higher values of the drag parameter $\mathcal{R}$ would be inconsistent with the observations.

\section{Discussion}
\label{sec:discuss}

The very low effective drag, $\mathcal{R}<10^{-2}$, implied by our analysis of the neutron star free precession in Her X-1, carries important consequences for the standard picture of superfluid dynamics and glitch phenomenology. In conventional glitch models, rapid spin-up events are attributed to the sudden unpinning and redistribution of angular momentum stored in a strongly pinned superfluid component, with subsequent coupling to the crust mediated by finite mutual friction and vortex creep. Unpinning occurs once the lag developed during the spin-down of the crust approaches the critical lag,
\begin{equation}
    \Delta \Omega_\mathrm{crit} \simeq \frac{f_\mathrm{p}}{\rho_\mathrm{s}\kappa\, r_{\rm NS}} \sim 0.1-1\quad \mbox{rad s}^{-1},
\end{equation}
where $f_\mathrm{p}$ is the pinning force per unit length, $\rho_\mathrm{s}$ the superfluid density, and $\kappa$ the quantum of circulation. This is larger by several orders of magnitude than the maximum lag developed during the quasi-steady precession phase (cf. Eq. \ref{eq:tilt_eq}):
\begin{equation}
    |\mathbf{\Omega}_\mathrm{s} - \mathbf{\Omega}_\mathrm{n}| \lesssim \Omega_\mathrm{n} \epsilon \approx 10^{-6} \quad \text{rad s}^{-1}.
\end{equation}
Consequently, the relative motion between the superfluid and the normal component induced by precession is far too small to trigger vortex unpinning.
In fact, the projected fractional lag obtained from Eq. \eqref{eq:lag_eq} for the net torque inferred in Her X-1, 
\begin{equation}
    \frac{|\delta \Omega_\mathrm{||eq}|}{\Omega_\mathrm{n}}\lesssim 10^{-9},   
\end{equation}
is several orders of magnitude smaller than the fractional spin jumps observed in glitches in many pulsars. This suggests that the 
dynamical state realized during long-lived precession is incompatible with the buildup of the large differential rotation required for 
glitch triggering. If the superfluid responsible for storing angular momentum in glitching pulsars behaves similarly in accreting 
systems such as Her X-1, the results imply that either the regions involved in glitches are largely decoupled from the precessing 
component, or that the effective vortex drag and pinning conditions depend sensitively on the dynamical state of the star.

Alternatively, the 35-day cycle may arise from a different mechanism, and the IXPE observations may admit an alternative interpretation. 
A low-mass third body orbiting the system has been proposed to explain the light curve \citep{Brecher1974, Mazeh1977}. However, this 
model predicts eccentricity of a few percents for the main binary orbit, which has been ruled out by accurate pulse-timing observations 
\citep{Deeter1981}. It is also difficult to see how such a model could account for the IXPE data.

\cite{Wolff1978} proposed an alternative scenario in which the periodicities in the light curve of Her X-1 are associated with 
oscillatory modes (g-modes) of the companion star, HZ Her. However, this model predicts additional periodicities on a timescale of $\sim 
1$ day, which are not observed. Moreover, it does not account for the periodic changes in the X-ray pulse profile that track the 35-day 
cycle.

The most viable alternative is therefore the precession of a warped, inclined accretion disc, conceivably driven by tidal torques from 
the companion star \citep{katz1973, Gerend1976, Petterson1977, Scott2000, Leahy2020, Leahy2025}. The modulation of the polarization 
signal could then arise from scattering within the accretion-disc corona (\citealt{Leahy2025}, but see \citealt{Heyl24}) although 
whether the IXPE data can be fitted with such a model remains to be demonstrated.  

\section{Conclusions}

We have examined the implications of long-lived free precession of the neutron star in Her X-1 for the dynamics of the internal 
superfluid. Sustaining precession over the $\sim$50 yr observational baseline requires an extremely weak effective coupling between the 
superfluid and the normal component, corresponding to a drag parameter $\mathcal{R}\lesssim10^{-2}$. 

This result carries important consequences for standard models of pulsar glitches. In particular, the differential rotation that develops during the quasi-steady precession phase is many orders of magnitude smaller than the critical lag expected for vortex unpinning, and is also far below the fractional spin changes observed in glitches. If the superfluid responsible for storing angular momentum in glitching pulsars behaves similarly in accreting neutron stars such as Her X-1, our results suggest that either the region involved in glitches is largely decoupled from the precessing component, or the effective pinning and drag depend sensitively on the dynamical state of the star.

Finally, we note that alternative explanations for the 35-day cycle in Her X-1 remain possible, and further observational and theoretical work will be required to determine whether they can account for the observed polarization properties.

\begin{acknowledgments}
   AL thanks Jeremy Heyl, Yuri Levin, Xin Sheng and Ira Wasserman for enlightening discussions, and the Center for Computational Astrophysics 
   for their warm hospitality and support. 
   This work was supported by a grant from the Simons Foundation (00001470).
\end{acknowledgments}


\appendix

\section{Effect of the crust-core coupling}
\label{app:core_effect}
\begin{figure*}
    \centering
    \includegraphics[width=\textwidth]{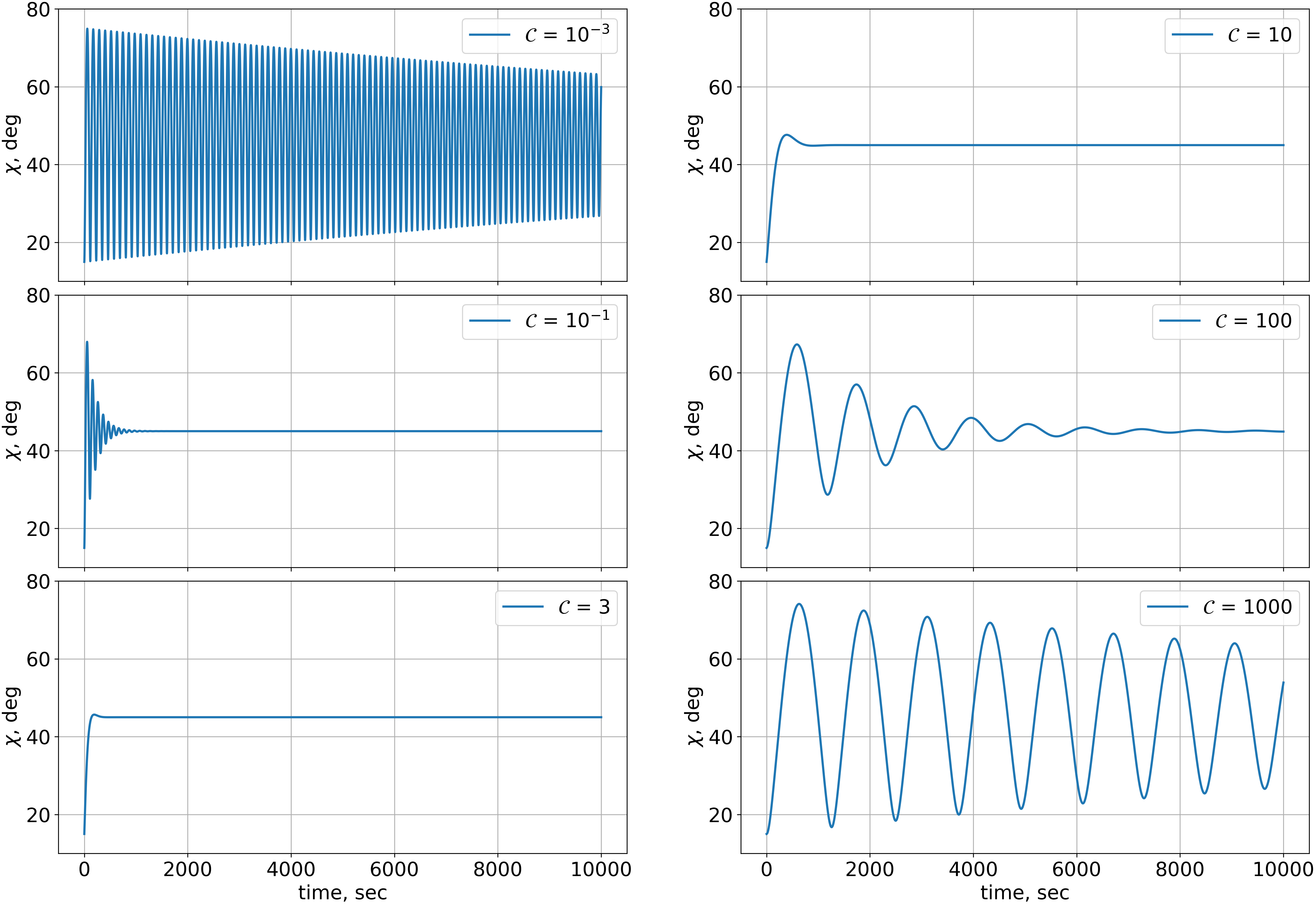}
    \caption{Evolution of the magnetic angle in a toy model including a strongly deformed crust ($\epsilon_\mathrm{cr} = 10^{-2}$) and a massive core ($\kappa_\mathrm{co} = 10$) coupled to it. The spin period of the star is $P_0 = 1\,\mathrm{s}$. The rotational evolution for different values of ${\cal C}$ is shown and is fully consistent with the theoretical predictions.  For weak coupling (${\cal C} = 10^{-3}$), the star exhibits long-lived precession with period $\sim 2P_0/\epsilon_\mathrm{cr} \sim 200\,\mathrm{s}$, for an initial tilt angle $\theta_0 = 60^\circ$, and with the magnetic axis inclined by $45^\circ$ with respect to the symmetry axis.
    For strong coupling (${\cal C} = 10^{3}$), the precession period is longer by a factor $\kappa_\mathrm{co}$, reaching $\approx 2000\,\mathrm{s}$. In the intermediate regime, precession damps rapidly, particularly near ${\cal C} = 3 \approx \sqrt{1 + \kappa_\mathrm{co}}$.}   
    \label{fig:core_chi}
\end{figure*}
Within the main body of the paper, we implicitly assume that the star's crust is tightly coupled to the core. Thus $I_0$ is the sum of the crust and core's moments of inertia: $I_0 = I_\mathrm{cr} + I_\mathrm{co} = I_\mathrm{cr}(1 + \kappa_\mathrm{co})$, where $\kappa_\mathrm{co} \sim 10^2$ for neutron stars. However, in reality, the crust-core coupling, even being strong, is not perfectly tight. Following  \cite{Alpar87}, one may assume that the crust is subject to the torque from the core
\begin{equation}
    \pmb N_\mathrm{co} = I_\mathrm{co}\tau^{-1}_\mathrm{co}(\pmb \Omega_\mathrm{co} - \pmb\Omega_\mathrm{n}),
\end{equation}
where $\pmb\Omega_\mathrm{co}$ is the core angular velocity and $\tau_\mathrm{co}$ is the coupling time scale. 

Consider the effect of imperfect crust–core coupling within a toy rotational model that neglects both the crustal superfluid and any external torques. In particular, we let the crust evolve according to
\begin{equation}
    \pmb {\dot \Omega}_\mathrm{n} + \pmb{\dot \omega} = (\pmb \omega \times \pmb \Omega_\mathrm{n}) + \kappa_\mathrm{co}\tau^{-1}_\mathrm{co}(\pmb \Omega_\mathrm{co} - \pmb\Omega_\mathrm{n}),
    \label{eq:crust_core_evol}
\end{equation}
where $\pmb\omega = (0, 0, \epsilon_\mathrm{cr}\Omega_\mathrm{n}\cos\theta)^T$, with $\epsilon_\mathrm{cr} = \Delta I_\mathrm{cr}/I_\mathrm{cr}$ characterizing the crust deformation.  Note that \(\epsilon_\mathrm{cr}\) does not represent the asymmetry of the total stellar moment of inertia, unlike the parameter \(\epsilon\) used in the main body of the paper. As will become clear later, \(\epsilon_\mathrm{cr}\) can be significantly larger than \(\epsilon\).

The rotational evolution of the core is given by
\begin{equation}
    \pmb{\dot \Omega}_\mathrm{co} = (\pmb{\Omega}_\mathrm{co} \times \pmb \Omega_\mathrm{n}) - \tau^{-1}_\mathrm{co}(\pmb \Omega_\mathrm{co} - \pmb\Omega_\mathrm{n}),
    \label{eq:core_evol}
\end{equation}
in the frame pinned to the crust.  Let
\begin{equation}
    {\cal C} = \dfrac{1 + \kappa_\mathrm{co}}{\Omega\tau_\mathrm{co}}
\end{equation}
be the dimensionless coupling parameter. In the limit ${\cal C} \rightarrow \infty$, the crust and core are perfectly coupled and rotate as a single rigid body. Conversely, for ${\cal C} = 0$, the core dynamics has no influence on the crust.
Since the second term on the right-hand side of Eqs.~(\ref{eq:crust_core_evol})--(\ref{eq:core_evol}) is dissipative, it plays a role analogous to that of the crustal superfluid discussed in the previous sections. Specifically, on a short initial timescale ($\sim \tau_\mathrm{co}/\kappa_\mathrm{co}$), the crust and core spin axes relax toward a quasi-equilibrium misalignment angle
\begin{equation}
    \gamma_\mathrm{cr-co,eq} = \epsilon_\mathrm{cr} \dfrac{\sin\theta\cos\theta}{\sqrt{1 + {\cal C}^2}}.
\end{equation}
After this transient stage, both components evolve jointly.

By rewriting Eqs.~(\ref{eq:crust_core_evol})--(\ref{eq:core_evol}) and assuming $\gamma_\mathrm{cr-co} \ll 1$, one finds that in the equilibrium state the crust precesses about the symmetry axis with angular velocity
\begin{equation}
    \dfrac{\mathrm{d}\varphi}{\mathrm{d}t} =
    \omega_\mathrm{cr} \dfrac{1 + (1 + \kappa_\mathrm{co})^{-1}{\cal C}^2}{1 + {\cal C}^2},
\end{equation}
while the tilt angle decays at the rate
\begin{equation}
    \dfrac{\mathrm{d}\theta}{\mathrm{d}t} =
    -\epsilon_\mathrm{cr}\kappa_\mathrm{co}\tau^{-1}_\mathrm{co}
    \dfrac{\sin\theta\cos\theta}{1 + {\cal C}^2}.
\end{equation}
Equivalently, in terms of the number of precession cycles,
\begin{equation}
    \dfrac{\mathrm{d} \theta}{\dot\varphi\, \mathrm{d}t}
    = - \dfrac{\kappa_\mathrm{co}{\cal C}}{1 + \kappa_\mathrm{co} + {\cal C}^2}\sin\theta,
\end{equation}
which is independent of $\epsilon_\mathrm{cr}$.

This result implies that in the no-coupling regime (${\cal C} = 0$), the crust precesses at the rate $\omega_\mathrm{cr}$ without any 
damping. This is a trivial result. On the other hand, in the limit of absolute coupling (${\cal C} \rightarrow \infty$), the precession 
becomes significantly slower:
\begin{equation}
    \dot\varphi = \frac{\omega_\mathrm{cr}}{1 + \kappa_\mathrm{co}},
\end{equation}
as expected, since in this case the star behaves as a single rigid body with total moment of inertia $I_\mathrm{cr}(1 + \kappa_\mathrm{co})$. In this limit, the precession is also undamped, $\dot\theta = 0$.

However, for finite coupling, the precession period exceeds $2\pi/\omega_\mathrm{cr}$, and the spin axis relaxes toward either the 
symmetry axis ($\epsilon_\mathrm{cr} > 0$) or the symmetry equator ($\epsilon_\mathrm{cr} < 0$) within a finite number of cycles.
The fastest damping occurs for ${\cal C} = \sqrt{1 + \kappa_\mathrm{co}}$, corresponding to
\begin{equation}
    \tau_\mathrm{co} = \frac{P\sqrt{1 + \kappa_\mathrm{co}}}{2\pi\kappa_\mathrm{co}} \approx \frac{P}{63},
\end{equation}
for $\kappa_\mathrm{co} = 10^2$. However, \cite{Alpar87} estimate $\Omega\tau_\mathrm{co} \sim 10^2 - 10^4$ for realistic neutron stars, which implies ${\cal C} \sim 10^{-2} - 1$. For the upper end of this range, the crust–core coupling damps the precession on a timescale of order one cycle. In contrast, if ${\cal C} \sim 10^{-2}$, the precession can survive for hundreds of cycles with period $\approx 2\pi/\omega_\mathrm{cr}$.

To illustrate the effect of crust–core coupling, we obtain a full numerical solution of the system (\ref{eq:crust_core_evol})--(\ref{eq:core_evol}) for a biaxial star with the following parameters: $\epsilon_\mathrm{cr} = 10^{-2}$, $P_0 = 1\,\mathrm{s}$, and $\theta_0 = 60^\circ$. The magnetic axis is assumed to be tilted by $45^\circ$ with respect to the symmetry axis, corresponding to an initial magnetic inclination $\chi_0 = 15^\circ$. The ratio of the core to the crust moments of inertia in this toy model is set to $\kappa_\mathrm{co} = 10$.

In Fig.~\ref{fig:core_chi} we show the simulated evolution of the magnetic angle in this model for a range of ${\cal C}$ values from $10^{-3}$ to $10^{3}$. The results are fully consistent with the theoretical predictions. For very weak coupling (${\cal C} = 10^{-3}$), the system exhibits long-lived precession with period $2\pi/\omega_\mathrm{cr} \approx 200\,\mathrm{s}$. However, a slow decay of the precession amplitude is observed.
For larger values of ${\cal C}$, damping of the precession becomes more significant, reaching its strongest regime near ${\cal C} \approx 3 \sim \sqrt{1 + \kappa_\mathrm{co}}$. Beyond this regime, strong coupling leads again to long-lived precession, but with a reduced frequency $\omega_\mathrm{cr}/(1 + \kappa_\mathrm{co})$.

\section{Steady-state rotation parameters}
\label{app:steady}

The evolution of $\delta\pmb\Omega = \pmb\Omega_\mathrm{s} - \pmb\Omega_\mathrm{n}$ in the linear approximation, Eq. \eqref{eq:Delta_evolution}, reads:
\begin{equation}
    \dfrac{\diff{(\delta\pmb \Omega)}}{\diff t} = -(\pmb\omega \times\pmb\Omega_\mathrm{n}) -\dfrac{\pmb N_\mathrm{ext}}{I_0} - (\pmb\Omega_\mathrm{Sh} \times \delta\pmb \Omega) - \dfrac{\delta\pmb \Omega +(\Omega_\mathrm{s}/\Omega_\mathrm{n}-1)\pmb \Omega_\mathrm{n}}{\tau_\mathrm{d}}.
    \label{eq:app:Delta_evolution}
\end{equation}
It is convenient to transform to the basis vectors $\pmb{e}_\pm = \pmb{e}_\varphi\pm i\pmb{e}_\theta$.  Then $\pmb{e}_\Omega\times\pmb{e}_\pm = \pm i\pmb{e}_\pm$,
and $\delta \pmb{\Omega}= \delta {\cal W}_\perp^-\pmb e_+ + \delta {\cal W}_\perp^+\pmb e_-$, where 
$2\delta{\cal W}_\perp^\pm = \Omega_\mathrm{n}(\varphi_\mathrm{s} - \varphi)\sin\theta \pm i\Omega_\mathrm{n}(\theta_\mathrm{s}-\theta)$.
For the parallel component, Eq. (\ref{eq:app:Delta_evolution}) reduces to
\begin{equation}
     \dfrac{\diff}{\diff t}(\delta{\Omega}_\parallel) + 2\tau_\mathrm{d}^{-1}\delta\Omega_\parallel  
    =-\Omega^2_\mathrm{n}\,n_\Omega,
    \label{eq:par_evol}
\end{equation}
and for the projection of the orthogonal component on $\pmb{e}_{\mp}$, to
\begin{equation}
    \dfrac{\diff}{\diff t} (\delta{\cal W}^{\pm}_\perp) + {\cal W}_\mathrm{Sh}\delta{\cal W}^{\pm}_\perp = -\Omega_\mathrm{n}^2 \dfrac{{\cal N}^{\pm}_\perp}{2},
    \label{eq:perp_evol}
\end{equation}
where ${\cal W}_\mathrm{Sh} \equiv \tau_\mathrm{d}^{-1} - i\Omega_\mathrm{Sh}$ and ${\cal N}^{\pm}_\perp \equiv 
(\epsilon\sin\theta\cos\theta+n_\varphi)\pm in_\theta$ are complex representations of the Shaham's frequency, and the projections of the 
torque on $\pmb{e}_{\mp}$, respectively. The dimensionless torque components $n_\theta$ and 
$n_\varphi$ are defined by Eq. (\ref{eq:dimless_torques}) above. Thus, $\tau_\mathrm{d} = (1 + {\cal R}^2)/(\Omega_\mathrm{n}{\cal R}(1 + \kappa_\mathrm{s}))$ is simply the decay time of the rapid Shaham's mode.

The external torque components change on a time-scale given by the minimum of two: the precession damping time $\tau_\mathrm{p}$ and the 
spin-down time. However, both are longer than $\tau_\mathrm{d}$ provided that $\kappa_\mathrm{s}/|\epsilon|\ll 1$. Therefore, the right-
hand sides of Eqs.~(\ref{eq:par_evol}) and (\ref{eq:perp_evol}) can be treated as constants evolving adiabatically. Their solutions are
\begin{equation}
    \delta\Omega_\parallel(t)=-\dfrac{1}{2}\Omega_\mathrm{n}^2n_\Omega\tau_\mathrm{d}\left(1 -e^{-t/2\tau_\mathrm{d}} \right) +\delta\Omega_{\parallel,0}\,e^{-t/2\tau_\mathrm{d}},
\end{equation}
and
\begin{equation}
    \delta{\cal W}^{\pm}_\perp(t)=-\Omega_\mathrm{n}^2\dfrac{{\cal N}^{\pm}_\perp{\cal W}_\mathrm{Sh}^*}{2|{\cal W}_\mathrm{Sh}|^2}\left(1 -e^{-{\cal W}_\mathrm{Sh}t}\right) 
    +\delta{\cal W}^{\pm}_{\perp,0}\,e^{-{\cal W}_\mathrm{Sh}t}
\end{equation}
respectively. These solutions show that if $\tau_\mathrm{d}$ is sufficiently short, the exponential terms decay rapidly, and the components of the lag approach slowly-evolving values. Consequently, both $\delta{\dot\Omega}_\parallel$ and $\delta\dot{\cal W}^{\pm}_\perp$ become close to zero, corresponding to a quasi-steady-state evolution. The corresponding lag components are
\begin{equation}
    \delta\Omega_{\parallel,\mathrm{eq}}=-\Omega_\mathrm{n}^2n_\Omega\tau_\mathrm{d}/2,
\end{equation}
and 
\begin{equation}
    \gamma^2_\mathrm{eq} = \delta{\cal W}^+_{\perp,\mathrm{eq}}\delta{\cal W}^-_{\perp,\mathrm{eq}} = \Omega_\mathrm{n}^4 \dfrac{{\cal N}^{+}_\perp{\cal N}^{-}_\perp}{|{\cal W}_\mathrm{Sh}|^2}.
\end{equation}
These expressions coincide with those given earlier in Eqs.~(\ref{eq:lag_eq}) and (\ref{eq:tilt_eq}), respectively.
The components of the latter along the polar and azimuthal directions are
\begin{equation}
    \delta \theta_\mathrm{eq} = -\dfrac{1-\kappa_\mathrm{s}{\cal R}^2}{1+\kappa_\mathrm{s}^2{\cal R}^2}\,(n_\varphi+\epsilon\sin\theta\cos\theta) - \dfrac{(1 + \kappa_\mathrm{s}){\cal R}}{1 + \kappa_\mathrm{s}^2{\cal R}^2} n_\theta,
\end{equation}
and
\begin{equation}
    (\delta \varphi\sin\theta)_\mathrm{eq} = \dfrac{1-\kappa_\mathrm{s}{\cal R}^2}{1+\kappa_\mathrm{s}^2{\cal R}^2}\,n_\theta - \dfrac{(1 + \kappa_\mathrm{s}){\cal R}}{1 + \kappa_\mathrm{s}^2{\cal R}^2}\,(n_\varphi+\epsilon\sin\theta\cos\theta)
\end{equation}
respectively.

Following the same approach, the precessional evolution can be treated by introducing the complex quantity ${\cal W}^+_\perp=\Omega_\mathrm{n}\varphi\sin\theta+i\Omega_\mathrm{n}\theta$, so that
\begin{equation}
    \dfrac{\diff}{\diff t}{\cal W}^+_\perp = \Omega_\mathrm{n}^2 {\cal N}^{+}_\perp + \Omega_\mathrm{n}\dfrac{\kappa_\mathrm{s}{\cal R}}{1 + {\cal R}^2}(1 + i{\cal R})\,\delta{\cal W}^+_\perp.
\end{equation}
If in this equation $\delta{\cal W}^+_\perp = \delta{\cal W}^+_{\perp,\mathrm{eq}}$, the solution coincides with those for $\dot\varphi$ and $\dot\theta$ obtained in the main text, Eqs.~(\ref{eq:dot_phi}) and (\ref{eq:dot_theta}), respectively.

\section{Hercules X-1 external torque}
\label{app:HerX1_torque}

As noted in Sect.~\ref{sec:HerX1} above, the neutron star in the Her X-1 system is subject to the influence of an external torque, which can be considered as a sum of several components:
\begin{equation}\label{E:AppC:N}
  {\pmb N}_\mathrm{ext,HX1} = \pmb N_\mathrm{acc} + \pmb N_\mathrm{mag} + \pmb N_\mathrm{psr} + 
  \pmb N_\mathrm{warp} + \pmb N_\mathrm{prec}.
\end{equation}
In this Appendix, we provide details of these torques as they have been adopted in the numerical calculations. The first component in the equation above represents the accretion spin-up:
\begin{equation}
  \pmb N_\mathrm{acc} = \dot M\, f(\cos\xi) \sqrt{GM_\mathrm{NS} r_\mathrm{m}} \pmb d =  N_\mathrm{acc,0}\,f(\cos\xi)\pmb d,
\end{equation}
where $\dot M$ is the accretion rate, $M_\mathrm{NS}$ is the star mass, $\pmb d$ is the disc angular momentum axis ($|\pmb d|=1$),
\begin{equation}
  r_\mathrm{m} = \dfrac{1}{2} \left(\dfrac{\mu^4}{2GM_\mathrm{NS}\dot M^2}\right)^{1/7}
\end{equation}
is the inner magnetospheric radius and
\begin{equation}
  f(\cos\xi) = A(\eta,\alpha,\chi)(1 - \eta\cos^2\xi) 
\end{equation}
is the modulation factor depending on the instant angle between the star's magnetic $\pmb m$ and disc axes: $\xi = \cos^{-1}(\pmb m\cdot \pmb d)$. The normalizing coefficient $A(\eta,\alpha,\chi)$ depend on the magnetic inclination $\chi$, spin-disc inclination 
$\alpha = \cos^{-1}(\pmb e_\Omega \cdot \pmb d)$ and modulation parameter $\eta = 0..1$ \citep[see][for details]{Biryukov2021} as
\begin{equation}
    A = \left[1- \dfrac{\eta}{2}\left(\sin^2\chi\sin^2\alpha + 2\cos^2\chi\cos^2\alpha \right) \right]^{-1}.
\end{equation}
The second term in Eq.~(\ref{E:AppC:N}) describes the spin-down due to the interaction of the disc by the stellar magnetic field:
\begin{equation}
  \pmb N_\mathrm{mag} = -\dfrac{\mu^2}{3 r_\mathrm{c}^3} \pmb e_\Omega = N_\mathrm{mag}\,\pmb e_\Omega,
\end{equation}
where
\begin{equation}
  r_\mathrm{c} = \left(\dfrac{GM_\mathrm{NS}}{\Omega^2} \right)^{1/3}
\end{equation}
is the corotation radius.

The third term represents pulsar losses:
\begin{equation}
    \pmb N_\mathrm{psr} = N_\mathrm{psr,0}(1 + \sin^2\chi)\pmb e_\Omega + N_\mathrm{psr,0} \cos\chi ((\pmb m \times \pmb e_\Omega) \times \pmb e_\Omega) + \dfrac{N_\mathrm{psr,0}}{3} \left( \dfrac{r_\mathrm{LC}}{r_\mathrm{NS}} \right) \cos\chi (\pmb m \times \pmb e_\Omega),
    \label{eq:psr_torque}
\end{equation}
and depends on the instant direction of the crust spin axis, magnetic angle $\chi$, and scales as
$N_\mathrm{psr,0} = -\mu^2/r_\mathrm{LC}^3$, where $r_\mathrm{LC} = c/\Omega$ is the light cylinder radius \citep{Abolmasov2024}.
At this point, we ignore pulsar losses enhancement due to the extra opening of the magnetic field lines by the disc \citep{parfrey16}, 
but take into account the so-called ``anomalous'' part of the pulsar torque. The latter scales as 
$(r_\mathrm{LC}/r_\mathrm{NS})N_\mathrm{prs,0}$ and hence is much stronger than standard pulsar losses. For this component, we adopted 
the equation found by \cite{bz_anomal2014} in the assumption that the star's internal magnetic field is uniform and aligned with the 
magnetic axis.

The two final terms account for backreaction from the warping and precession of the tilted disc:
\begin{equation}
  \pmb N_\mathrm{warp} = \dfrac{4\sqrt{2}}{3} N_\mathrm{acc,0} \cos^2\chi\cos\alpha\, (\pmb d \times (\pmb e_\Omega \times \pmb d)) 
\end{equation}
and
\begin{equation}
  \pmb N_\mathrm{prec} = \dfrac{32\sqrt{2}}{3\pi} N_\mathrm{acc,0} \sin^2\chi\cos\alpha\, (\pmb e_\Omega \times \pmb d)
\end{equation}
respectively \citep{Lai99, 2011MNRAS.412.2790L}. 

Until the disc angular momentum is aligned to the star spin axis ($\alpha = 0$), the fundamental equations (\ref{eq:Euler_crust}) and (\ref{eq:Euler_superfluid}) must be supplemented with another one, describing the evolution of $\alpha$:
\begin{equation}
    I_0\Omega \dfrac{\diff \alpha}{\diff t}= - {\pmb N}_\mathrm{ext,HX1}\cdot\pmb l,
    \label{eq:dot_alpha_general}
\end{equation}
where $\pmb l$ is the unit vector that always lies within the $\pmb e_\Omega -\pmb d$ plane, orthogonal to the
spin axis, and makes an acute angle with $\pmb d$. However, $\pmb N_\mathrm{ext,HX1}$ includes rapidly oscillating torques ${\pmb N}_\mathrm{acc}$, ${\pmb N}_\mathrm{warp}$, and  ${\pmb N}_\mathrm{prec}$ that rotate around $\pmb \Omega$ with on the spin period time scale, when considered in the principal frame. Thus, following the approach used by \cite{Biryukov2021}, we hereafter average the net torque projections over the spin period to obtain analytic expressions for the components of ${\pmb N}_\mathrm{ext,HX1}$. In particular Eq.(\ref{eq:dot_alpha_general}) then reads as
\begin{equation}
    \left \langle I_0 \Omega \dfrac{\diff \alpha}{\diff t} \right\rangle = -\left \langle {\pmb N}_\mathrm{ext,HX1}\cdot {\pmb l} \right\rangle = - N_\mathrm{acc,0}\sin\alpha\,\left(1 - \dfrac{4\sqrt{2}}{3}\cos^2\alpha\cos^2\chi \right).
\end{equation}
Since the expression in the parentheses on the RHS of this equation is capable of changing sign, a stationary value of $\alpha$ can exist, in principle. Spin-disk alignment reaches it on the accretion time scale and then continues to evolve on the $|\chi/\dot\chi|$ time scale (see Fig. 4 from \cite{2011MNRAS.412.2790L}). The detailed analysis of that is beyond the scope of our paper.



\bibliography{sample701}{}
\bibliographystyle{aasjournalv7}



\end{document}